\documentclass{aa}  

\usepackage{graphicx}
\usepackage{txfonts}
\usepackage{ulem}
\usepackage{color}
\begin{document}

\title{ Electron acceleration driven by the lower-hybrid-drift instability}
\subtitle{An extended quasilinear model}

\author{ Federico Lavorenti
          \inst{1,2}\thanks{federico.lavorenti@oca.eu}
          ,
         Pierre Henri
          \inst{1,3}
          ,
         Francesco Califano
          \inst{2}
          ,
         Sae Aizawa
          \inst{4}
          and
         Nicolas Andr\'e
          \inst{4}
          }
          
 \institute{ Laboratoire Lagrance, Observatoire de la C\^{o}te d’Azur, Universit\'e C\^{o}te d’Azur, CNRS, Nice, France
 \and
 Dipartimento di Fisica "E. Fermi", Università di Pisa, Pisa, Italy
 \and
LPC2E, CNRS, Univ. d'Orl\'eans, OSUC, CNES, Orl\'eans, France
 \and
IRAP, CNRS-CNES-UPS, Toulouse, France
              }

\titlerunning{LHDI electron acceleration in space plasmas}

\authorrunning{Lavorenti et al.}

 \date{Submitted 11 April 2021}

\abstract
{Density inhomogeneities are ubiquitous in space and astrophysical plasmas, in particular at contact boundaries between different media. They often correspond to regions that exhibits strong dynamics on a wide range of spatial and temporal scales. Indeed, density inhomogeneities are a source of free energy that can drive various instabilities such as, for instance, the lower-hybrid-drift instability which in turn transfers energy to the particles through wave-particle interactions and eventually heat the plasma. }
{We aim at quantifying the efficiency of the lower-hybrid-drift instability to accelerate and/or heat electrons parallel to the ambient magnetic field.}
{We combine two complementary methods: full-kinetic and quasilinear models.} 
{We report self-consistent evidence of electron acceleration driven by the development of the lower-hybrid-drift instability using 3D-3V full-kinetic numerical simulations.
The efficiency of the observed acceleration cannot be explained by standard quasilinear theory.
For this reason, we develop an extended quasilinear model able to quantitatively predict the interaction between lower-hybrid fluctuations and electrons on long time scales, now in agreement with full-kinetic simulations results.
Finally, we apply this new, extended quasilinear model to a specific inhomogeneous space plasma boundary: the magnetopause of Mercury, and we discuss our quantitative predictions of electron acceleration in support to future BepiColombo observations.}
{}


\maketitle

\section{Introduction}

Inhomogeneities in magnetic field, velocity, density, temperature, etc. from fluid down to kinetic scales are commonly encountered in space and astrophysical plasmas~\citep{Amatucci1999}. The gradient associated to such inhomogeneous plasma regions is the source of "free" energy that can drive various kind of plasma instabilities.
For instance, in the case of density gradient on scale close to the ion gyroradius, a situation commonly encountered in many plasma environments, the plasma is unstable against the so-called drift instabilities.
These instabilities arise from the relative motion between ions and electrons, and turn out to be of paramount importance in shaping plasma boundaries found in space, allowing for strong anomalous mass and energy transport not achievable by standard collision-like diffusion.

\textit{In situ} measurements show that lower-hybrid waves (hereafter LHW) with frequency close to the lower-hybrid frequency $f_{_{LH}}\approx\sqrt{\omega_{ci}\omega_{ce}}$ are ubiquitous in magnetized space plasma environments.
Such waves are commonly observed at Earth's magnetotail~\citep{Huba1978,Retino2008,Zhou2009,Zhou2014,Khotyaintsev2011,Norgren2012,LeContel2017} and Earth's magnetopause~\citep{Andre2001,Bale2002,Vaivads2004,Graham2017,Graham2019,Tang2020}.
In these two regions LHW are commonly observed in the vicinity of magnetic reconnection sites where strong density gradients do form. Their role on the onset (and/or relaxation) of magnetic reconnection has been addressed in the past and still represents a key point in the context of reconnection research~\citep{daughton2003,Lapenta2003,Lapenta2018,Yoo2020}.

Moreover, LHW are also observed at plasma shock fronts such as the terrestrial bow shock~\citep{Walker2008}, interplanetary shocks in the solar wind~\citep{Volodia1985,Zhang1998,Wilson2013}, and supernova remnants~\citep{Laming2001}. 
Finally, LHW have been observed in induced ionosphere of comet 67P~\citep{Andre2017,Karlsson2017,Goldstein2019}, of the planets Venus~\citep{Scarf1980,Shapiro1995}, Mars~\citep{Sagdeev1990} and at Earth's ionosphere~\citep{Reiniusson2006}.
In this context, space observations of supra-thermal electron populations in conjunction with LHW represents one of the basic points motivating the interest in the study of the interaction of these waves with electrons~\citep{Norgren2012,Zhou2014,LeContel2017,Broiles2016,Goldstein2019}.

More than being just a mechanism at work in space plasma environments, electron acceleration by LHW is a mechanism commonly used in tokamak experiments to heat electrons and hence the plasma along the toroidal magnetic field lines~\citep{Becoulet2011,Pericoli-Ridolfini1999}. This naturally suggests that LHW generated in space are an efficient driver for electron acceleration in space plasma environments~\citep{Broiles2016}. However, the mechanism for the generation of LHW in laboratory differs much from the one at play in space plasmas. In plasma fusion experiments the LHW are commonly excited by an external pump 
enabling to sustain the waves and so the electron acceleration process on long time scales with parameters controlled by the experimenter himself. In inhomogeneous space plasma instead LHW are generated by the development of plasma instabilities, which nonlinear saturation might reduce the efficiency of the acceleration process when compared to plasma fusion experiments. 

In a natural plasma environment the two instabilities at play for the generation of LHW are (i) the modified-two-stream instability (hereafter MTSI) driven by a supra-thermal ion beam~\citep{Ott1972,McBride1972,McBride1972_beta} and (ii) the lower-hybrid-drift instability (hereafter LHDI) driven by the relative drift between ions and electrons~\citep{Krall1971,Krall1973,Gary1993}. 
The electron acceleration driven by LHW generated by the MTSI has been widely addressed in the literature using quasilinear theory \citep{McBride1972,Shapiro1999}, full-kinetic simulations \citep{McClements1993,bingham2002} and experiments \citep{rigby2018}. 
The MTSI is considered to be the typical source for the above mentioned LHW observations at plasma shock fronts~\citep{Volodia1985,Shapiro1995,Shapiro1999} due to the reflection of a large fraction of solar wind ions by the shock front; however, LHW are routinely observed also in the absence of such beams of reflected ions, a condition that prevents to invoke the MTSI as the underlying mechanism.
In such cases, the LHW are instead generated by the LHDI. The driver for the development of the LHDI has to be found in the ``strong'' density gradients reaching length scales of the order of, or even shorter than, the ion gyro-radius. This is the case, for instance, for the above-mentioned observations at Earth's magnetosphere and in cometary plasmas.
From now on, we shall focus on LHW generated by the LHDI and their interaction with the electron population.

The LHDI fastest growing modes propagate perpendicular to both the density gradient, let say the $x$-direction, and the ambient magnetic field direction, the $z$-direction. The phase velocity is of the order of the ion thermal speed~\citep{Gary1993}. 
However, the LHDI modes are unstable over a narrow cone angle around this direction (in this case the $y$-direction) proportional to the square root of the ion-to-electron mass ratio $k_z/k_y\approx\sqrt{m_e/m_i}$~\citep{Gary1978}. 
This means that the oblique LHDI modes have a component of the phase velocity parallel (resp. perpendicular) to the ambient magnetic field of the order of the electron (resp. ion) thermal speed. As a result, electrons can be resonantly accelerated by LHDI fluctuations in the direction parallel to the ambient magnetic field through wave-particle interactions, mechanism hereafter called LHDI electron acceleration.

The goal of this paper is to investigate the efficiency of the LHDI electron acceleration.
As of today, the state-of-the-art on this mechanism is provided by the analytical quasilinear model proposed by~\cite{Cairns2005}, hereafter called the QL model. Such QL model is well suited to study the early stage of the electron acceleration, but eventually it breaks down when the nonlinear feedback from the particles distribution to the wave becomes important. More precisely the QL model breaks down as soon as nonlinear effects locally modify the distribution function shape (i.e. around the resonant velocity), hereafter called nonlinear LD-like effects because analogous to the well known nonlinear Landau damping effects (see~\citet{Brunetti2000} and references therein).

Other past works have addressed the problem of the interaction between pump-generated LHW and electrons including such nonlinear LD-like effects~\citep{Singh1996,Singh1998,Zacharegkas2016}. However, the configuration adopted in these works is not well suited for space plasma configurations where LHW are typically driven by a plasma instability, like the LHDI, and not by an external pump.
To our best knowledge, the LHDI electron acceleration mechanism has not yet been studied using a self-consistent full-kinetic model. 

In the past, the efficiency of LHDI electron acceleration has been addressed by means of reduced analytical models like the QL model, mostly because of the limitations on computational resources. Indeed, the computational power required to solve the Vlasov equation for LHDI electron acceleration is challenging because (i) ion and electron kinetic physics must be included self-consistently in the model, meaning that a full-kinetic numerical simulation is required and (ii) the wave propagation occurs over an angle, in the $y-z$ plane, perpendicular to the inhomogenity direction, the $x$-direction, meaning that three-dimensional numerical simulations are required.
All in all, investigating the electron acceleration generated by the LHDI requires full-kinetic 3D-3V simulations.
This is one of the methods used in this work.

In this paper we investigate the electron acceleration associated to the LHDI through a comparison of complementary numerical simulations. We use the quasilinear approach and direct full-kinetic 3D-3V simulations. This enables us to assess the intrinsic limits of the QL model and to investigate the consequences of  nonlinear LD-like effects on the LHDI electron acceleration.
We present first direct numerical evidence of LHDI electron acceleration from full-kinetic 3D-3V simulations, and we build up an extended quasilinear (eQL) model that takes into account the effect of such nonlinear saturation to quantitatively estimate electron acceleration under realistic space plasma parameters.

The paper is organised as follows: 
Sec.~\ref{sec:models} describes the QL model and the full-kinetic 3D-3V simulation model we use here. 
Sec.~\ref{sec:results} presents the results of both models using two common sets of plasma parameters (``strong'' and ``weak'' gradient configurations).
Sec.~\ref{sec:discussion} compares the results of the two models, shows the limitations of the QL model as compared to the full-kinetic one, and presents a novel eQL model. 
Finally, we apply this new eQL model to the magnetopause of Mercury in view of the future observations of the BepiColombo space mission. 
Sec.~\ref{sec:conclusions} summarizes and concludes this work.

\section{Models and methods}\label{sec:models}

In this study, we use two different models of plasma evolution. First, a QL model of LHDI electron interaction based on the work of~\citet{Cairns2005}.
Second, a full-kinetic 3D-3V plasma simulations of a plasma boundary initially unstable to the LHDI. 
The former is a simplified model of the plasma dynamics that does not account for the full response of the plasma itself to nonlinear interactions, and therefore is considered a \textit{reduced} model. The latter instead is fully self-consistent, even if constrained by a specific parameter choice, and therefore it is considered an \textit{ab initio} model.
The full-kinetic model, being more general than the QL one, is used to assess the limits of the latter and eventually, to build an extended description that properly models the LHDI electron acceleration still within a quasilinear framework.

\subsection{Quasilinear analytical model}\label{subs:ql_model}

The wave-particle interaction between LHW (generated from LHDI) and electrons can be modelled using a powerful analytical tool: quasilinear theory~\citep{Bernstein1966,Alexandrov1984}. Quasilinear theory is based on a second order perturbative expansion of the Vlasov equation averaged over the spatial variables.
The system of quasilinear equations describes (i) the diffusion in velocity space of the electron distribution function trough a diffusion coefficient proportional to the electric field energy (Eqs.~\ref{eq:QL_cairns_fe}-\ref{eq:QL_cairns_De}), and (ii) the time evolution of the electric field energy (Eqs.~\ref{eq:QL_cairns_Sk}-\ref{eq:QL_cairns_gammae}). 
The state-of-the-art QL model for LHDI electron interaction is the one developed in~\citet{Cairns2005} and summarized here:
\begin{eqnarray}
    \partial_t f_e(v_{\parallel},t) & = &\label{eq:QL_cairns_fe}
                \partial_{v_{\parallel}} D_e(v_{\parallel},t) \partial_{v_{\parallel}} f_e\\
    D_e(v_{\parallel},t)            & = &\label{eq:QL_cairns_De}
                \frac{e^2}{4 \epsilon_0 m^2_e} \int S_k(k_{\bot},k_{\parallel},t) \frac{k^2_{\parallel}}{k^2_{\bot}} \delta(\omega-k_{\parallel}v_{\parallel}) d^3\mathbf{k}\\
    \partial_t S_k(k_{\bot},k_{\parallel},t) & = &\label{eq:QL_cairns_Sk}
                \left[ \gamma_{_{LHDI}} \left( 1-\frac{S_k}{S_{k,max}} \right) +\gamma_e(k_{\bot},k_{\parallel},t) \right] S_k\\
    \gamma_e(k_{\bot},k_{\parallel},t) & = &\label{eq:QL_cairns_gammae} 
                \frac{\pi \omega^2_{_{LH}} \omega(k_{\bot},k_{\parallel})}{2 n_0 k^2_{\bot}} \frac{m_i}{m_e} \partial_{v_{\parallel}} f_e (v_{\parallel}=\omega/k_{\parallel},t)\,
\end{eqnarray}
There, $f_e(v_\parallel,t)$ is the electron distribution function, $k_{\bot}$ (resp. $k_{\parallel}$) is the wavevector perpendicular (resp. parallel) to the ambient magnetic field, $S_k=E^2_k/8\pi$ is the electric field energy density in wavevector-space, $S_{k,max}$ is the maximum value of $S_k$ attained at saturation, $n_0$ is the plasma density, $\omega_{_{LH}}$, $\omega_{ci}$ are the lower-hybrid and ion cyclotron frequencies, $\omega(k_{\bot},k_{\parallel})$ is the spectrum of the wave, $\gamma_{_{LHDI}}$ is twice the linear growth rate of the LHDI and $\delta(x)$ is the Dirac delta function.
In the following, $\rho_i$ is the ions gyroradius, and $v_{thi}=\rho_i\omega_{ci}$ is the ion thermal speed.

After normalization, this nonlinear system of coupled partial differential equations (Eqs.\ref{eq:QL_cairns_fe}-\ref{eq:QL_cairns_gammae}) is solved by numerical integration using a time staggered leapfrog scheme.
In $k$-space, we limit the computation to the region of LHDI fastest growing modes, i.e. $0.7<k_{\bot}\rho_e<1$ and $0<k_{\parallel}\rho_i<1$, as done in~\citet{Cairns2005}.
Since the wavevectors are limited to this region, the wave spectrum turns out to be more or less flat with a corresponding frequency for the fastest growing mode given by the lower-hybrid frequency,  $\omega(k_\bot,k_\parallel) \simeq \omega_{_{LH}}$.
The grid in velocity space is chosen by testing the convergence of the solution.

The QL model, Eqs.~(\ref{eq:QL_cairns_fe}-\ref{eq:QL_cairns_gammae}), depends on four parameters: $m_i/m_e$, $\omega_{pe}/\omega_{ce}$,  $\gamma_{_{LHDI}}$ and $S_{k,max}$. 
The first two parameters  $m_i/m_e$, $\omega_{pe}/\omega_{ce}$ define the plasma itself. 
The last two parameters $\gamma_{_{LHDI}}$ and $S_{k,max}$ (or analogously $S_{max}=\int d^3\mathbf{k} S_{k,max}$) define the linear growth and saturation level of the instability, and they only depend on the initial plasma configuration.
The analytical expressions for these two quantities are given by:
\begin{eqnarray}
    \gamma_{_{LHDI}} & = &\label{eq:gammaL_saturation}
        \frac{\sqrt{2\pi}}{4}\frac{1}{\sqrt{1+\beta_i/2}} (\epsilon_n \rho_i)^2 \omega_{_{LH}} \\
    S_{max}  & = &\label{eq:SnTi_saturation}
        \begin{cases}
        2\frac{m_e}{m_i} \frac{(\epsilon_n \rho_i)^2}{(1+\omega^2_{pe}/\omega^2_{ce})} n_0T_i & \text{current relaxation}\\
        \frac{2}{45\sqrt{\pi}} \frac{(\epsilon_n \rho_i)^5}{(1+\omega^2_{pe}/\omega^2_{ce})} n_0T_i   & \text{ion trapping}
         \end{cases}\,
\end{eqnarray}
where $\beta_i$ is the ion plasma beta, and the inverse gradient scale length $\epsilon_n$ is defined as 
\begin{equation}
\label{eq:epsilonn}
\epsilon_n = \max \left\{ \frac{1}{n(x)}\frac{d n(x)}{dx} \right\}
\end{equation}
The growth rate $\gamma_{_{LHDI}}$ in Eq.~(\ref{eq:gammaL_saturation}) has been obtained using a linearized kinetic model by~\citet{Davidson1977}.
The electric energy at saturation $S_{max}$ in Eq.~(\ref{eq:SnTi_saturation}) has been obtained using an analytic quasilinear approach by~\citet{Davidson1978}, and it has been later tested numerically by~\citet{Brackbill1984} using 2D full-kinetic simulations. 
The LHDI can saturate through two different processes depending on the initial value of the density gradient: ion trapping (resp. current relaxation) for high (resp. low) values of the density gradient (and therefore of the drift speed)~\citep{Brackbill1984}. 

The derivation of Eqs.~(\ref{eq:gammaL_saturation}-\ref{eq:SnTi_saturation}) is based on the assumption that the only source of drift in the plasma is the density gradient. Thus, all particle drifts, except the diamagnetic drift $v_{Di}$, are considered negligible.
As a consequence,  $v_{Di}/v_{thi}=\epsilon_n \rho_i$.
In the full-kinetic simulations, presented in the next Section, this assumption is substantially verified.

Quasilinear models are inherently limited because they do not include the nonlinear feedback from the modified plasma dispersion function on the electromagnetic fields.
To overcome this limitation, we present in the next Section a full-kinetic 3D-3V numerical plasma model.

\subsection{Setup full-kinetic 3D-3V simulations}\label{subs:fullkinetic_model}

The full-kinetic model of the plasma is based on a direct solution of the Vlasov-Maxwell system of equations using a Lagrangian PIC (particle-in-cell) approach. 

\begin{figure}
    \centering
    \includegraphics[width=\linewidth]{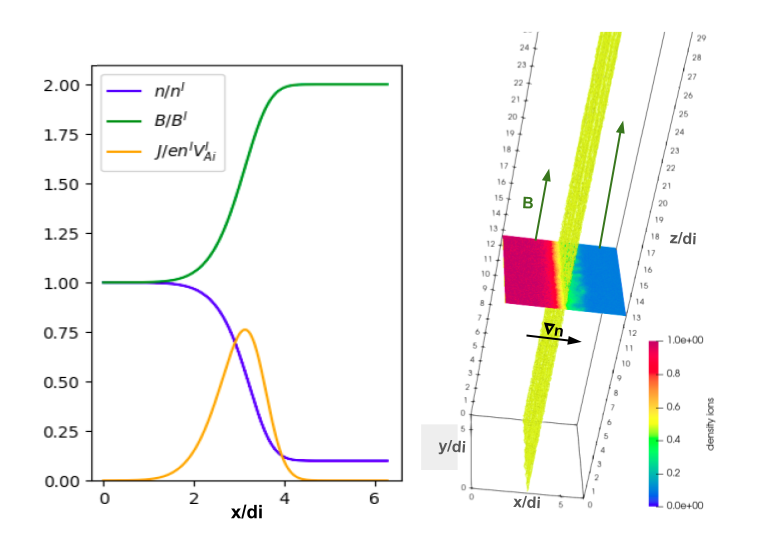}
    \caption{Left panel: density, magnetic field, and current profiles at $t=0$ along the direction of the inhomogeneity ($x$-axis) in the ``strong gradient'' simulation. Right panel: 3D visualization of the ion density in the ``strong gradient'' simulation at $t\omega^I_{ci}=12$, the LHDI fluctuations are highlighted in the unstable plane (yellow) and in a cut perpendicular to the ambient magnetic field.}
    \label{fig:initial_cond_pic}
\end{figure}

To run the simulations of a plasma boundary unstable to the LHDI, we have used the explicit, electromagnetic, relativistic, PIC code SMILEI~\citep{Derouillat2018}. The ambient magnetic field is directed along the $z$-axis and the density gradient along the $x$-axis. 
In order to model both wave propagation (predominantly along the $y$-axis, perpendicular to both the magnetic field and the density gradient direction) and the electrons wave-particle interaction (predominantly along the $z$-axis, parallel to the magnetic field), we consider a three-dimensional (3D) numerical box. 
Compared to previous numerical investigations of the LHDI \citep{Brackbill1984,Gary1990,Hoshino2001,Shinohara1999,Lapenta2002} that focused on the wave generation mechanism only through 2D-3V simulations in the equivalent of our ($x,y$) plane, we also include in this study the out-of-plane direction that hosts the electron acceleration resonant processes.
An overview of the numerical setup is shown in Fig.~\ref{fig:initial_cond_pic}, the right panel shows the 3D numerical box used here, highlighting (i) a slice of the ion density field in the ($x,y$) plane, and (ii) the plane ($y,z$) most unstable to the LHDI in yellow.

The simulations are initialized ensuring pressure balance and using the Vlasov equilibrium proposed in \citet{Alpers1969,Pu1981}, which shapes a plasma boundary with density and magnetic field asymmetries, uniform temperature, no electric field and no velocity nor magnetic field shear.
Hereafter, we shall refer to the side $I$ (resp. side $II$) of this boundary as the high (resp. low) density side.
The expressions for the initialization profiles are: 
\begin{eqnarray}
        n(x) & = &\label{eq:n_profile_kin}
        n^{I} + \frac{n^{II}-n^{I}}{2}\left( 1+ \text{erf} \left( w \frac{eA_y(x)}{m_i v_{thi}}  \right) \right)\\
        B_z(x) & = &\label{eq:Bz_profile_kin}
        \sqrt{8\pi} \left( PBC - n(x)T \right)^{1/2}\\
        B_z(x) & = &\label{eq:Ay_profile_kin}
        \frac{dA_y(x)}{dx}~~~~\text{with}~~A_y(x_0)=0
\end{eqnarray}
Where, $T=T_i+T_e$ is the uniform temperature, $PBC$ is the constant of pressure balance (set to have the plasma beta $\beta^I=10/3$ and $\beta^{II}=1/12$), and $w$ is a constant that defines the width of the layer -- set to 0.98 (resp. 0.5) in the ``strong'' (resp. ``weak'') gradient case. We use a density and magnetic field asymmetry of $n^I/n^{II}=10$ and $B^I/B^{II}=0.5$.
The shape of these initialization profiles is shown in the left panel of Fig.~\ref{fig:initial_cond_pic}.
In the following, all quantities are normalized to ion quantities in side $I$: the ion gyro-frequency $\omega^{I}_{ci} {=} eB^I/m_i c$, the ion skin depth $d^I_i {=}  c \sqrt{m_i/4\pi n^I e^2}$ and the Alfv\'{e}n speed $V^{I}_{A} {=} B^I/\sqrt{4\pi m_i n^I}$.
The numerical box dimensions are $L_x {=} L_y {=} 2\pi$ and $L_z {=} 20\pi$ with a number of cells $N_x {=} N_y {=} 288$ and $N_z {=} 1472$.
The box is elongated -- by a factor $\sqrt{m_i/m_e} {=} 10$ -- in the magnetic field direction in order to reliably include the narrow cone angle over which unstable LHDI modes grow. 
The timestep used in the simulations is $dt = 3.4 \times 10^{-4}$ to satisfy the CFL stability condition. 
The simulations are ran for several tens of the ion gyro-periods until electron acceleration saturates. 
We use 100 macro-particles per cell and a second order spline interpolation for the macro-particles. 
We use a reduced ion-to-electron mass ratio $m_i/m_e=100$ for the sake of computational reasons. The implications of using such a reduced mass ratio are discussed in Sec.~\ref{sec:discussion}. 
We use an ion-to-electron temperature ratio $T_i/T_e=1$, and a plasma-to-cyclotron frequency ratio of $\omega^I_{pe}/\omega^I_{ce}=4$.
The magnitude of the density asymmetry used in these simulations encompasses the typical parameters observed in small planetary magnetospheres such as Mercury~\citep{Gershman2015}.

Two different simulations are investigated using two different layer widths: (i) a steeper boundary case (with the inverse gradient scale length $\epsilon_n {=} 1$, defined in Eq.~\ref{eq:epsilonn}), hereafter called ``strong gradient'' simulation, and (ii) a smoother boundary case (with $\epsilon_n {=} 0.5$) hereafter called ``weak gradient'' simulation. 
These parameters are chosen, first, in order to ensure that the LHDI fluctuations amplitude are well above the PIC noise level, and, second, in order to saturate through the two different mechanisms introduced in Sec.~\ref{subs:ql_model}: ion trapping or current relaxation.

\section{Evolution of LHDI and associated electron acceleration: simulations results}\label{sec:results}

First, we show the results of the full-kinetic model.
Then, we show those obtained with the QL model using the same parameters as for the full-kinetic simulations.
Finally, we compare the results of the two models showing the range of validity and the limitations of the QL model.

\subsection{Results from the full-kinetic 3D-3V simulations}\label{subs:results_fullkinetic}

\begin{figure}
    \centering
    \includegraphics[width=\linewidth]{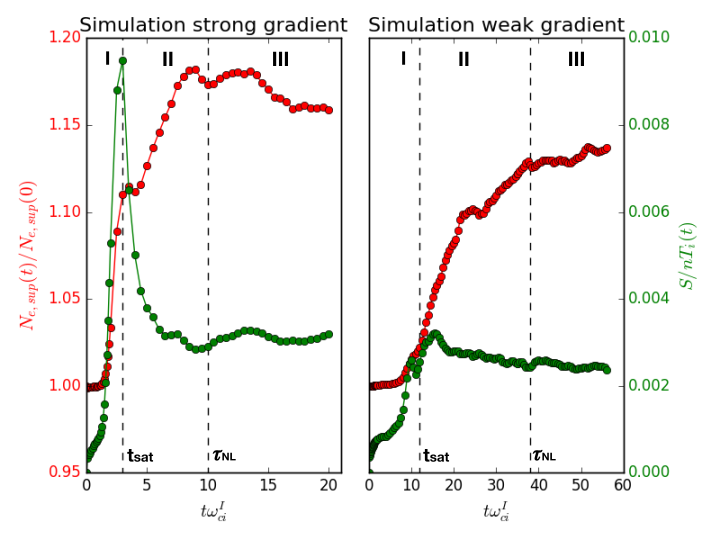}
    \caption{Full-kinetic simulations: evolution of the electron supra-thermal density $N_{e,sup}(t)/N_{e,sup}(0)$ (red curves) and of the electric field energy normalized to the ion thermal energy $S/nT_i(t)$ (green curves), both curves are integrated over the unstable layer ($3.5<x<4$ for strong gradient and $3.5<x<4.5$ for weak gradient). Note that time axes are different for the two simulations.}
    \label{fig:electron_acc}
\end{figure}

In both ``strong'' and ``weak'' gradient, full-kinetic simulations the layer is unstable to the LHDI due to the presence of a density gradient on ion kinetic scales. 
The LHDI fluctuations grow exponentially in the layer for times $t<t_{sat}$ as predicted by kinetic linear theory. The fastest growing mode (hereafter FGM) is electrostatic and directed along the $y$-axis with wavevector $k_y \rho_e \approx 1$, frequency $\omega \lessapprox \omega_{_{LH}}$ and a growth rate $\gamma \lessapprox \omega_{_{LH}}$ in agreement with the linear estimation (Eq.~\ref{eq:gammaL_saturation}) for both simulations. The growth of the electric field energy -- normalized to the ion thermal energy and integrated over the unstable layer -- is shown in Fig.~\ref{fig:electron_acc}, green curves, for both simulations.  
At $t\approx t_{sat}$ (corresponding to the first vertical dashed lines in each panel of Fig.~\ref{fig:electron_acc}), the electric field fluctuation's growth saturates.
Using the growth rate and the saturation level from Eqs.~(\ref{eq:gammaL_saturation}-\ref{eq:SnTi_saturation}), we compute the saturation time analytically as
\begin{equation}\label{eq:saturation_time}
    t_{sat}=\frac{\ln \left[ S_{max}/S(t=0) \right] }{\gamma_{_{LHDI}}}
\end{equation}
The saturation mechanism of the LHDI have been extensively studied in the literature in the past, see for instance~\citet{Brackbill1984,Chen1983,Davidson1978}. In the strong gradient simulation the LHDI is observed to saturate due to ion trapping, while in the weak gradient simulation, the LHDI saturates due to current relaxation, as expected.
In both cases, the saturation levels are comparable with the ones obtained by Eq.~(\ref{eq:SnTi_saturation}). 
Subsequently, in the strong gradient simulation, we observe for $t>t_{sat}$ a rapid decrease of the electric fluctuations amplitude, characteristic of an overshoot pattern, shown in the electric-to-thermal energy ratio in Fig.~\ref{fig:electron_acc} (left panel, green curve) at time $t=t_{sat}$.
Such overshoot of the electric field fluctuations is not observed in the weak gradient simulation.
An important point to mention is that, in both simulations, the electric field energy always remains much smaller than the thermal energy of the particles (it never exceeds 1\%). 

Note that a common issue with PIC simulations is the possible occurrence of spurious numerical heating of the macroparticles population during the simulation. 
This effect, due to the so-called ``finite grid instability''~\citep{Birdsall_Langdon1991,Markidis2010,McMillan2020}, appears when the Debye length is not resolved enough by the numerical grid. In such a case, the electrons would be numerically heated, which would hide the wave-particle interaction we are looking for. We have carefully checked that our PIC simulations are free from such numerical spurious heating, by comparing the evolution of the supra-thermal electron density in the unstable layer to the one outside the layer on both sides, where the plasma is stable, to check that the numerical heating of the electrons is negligible compared to the one arising from wave-particle interaction in the layer.

To quantify the efficiency of the LHDI in accelerating electrons parallel to the ambient magnetic field, we define a supra-thermal electrons density as follows:
\begin{equation}\label{eq:tracer}
    N_{e,sup}(t) = \int^{2v_{the}}_{-\infty} f_e(v_z,t) dv_z + \int^{\infty}_{2v_{the}} f_e(v_z,t) dv_z 
\end{equation}
In the following, the supra-thermal electrons density $N_{e,sup}$ is used as a quantitative tracer of the LHDI electron acceleration. 
The growth and saturation of $N_{e,sup}$ is shown for both simulations in Fig.~\ref{fig:electron_acc}, red curves.

In both simulations, the efficiency of the LHDI electron acceleration process can be described through a three-phase evolution: (i) the linear phase $0<t<t_{sat}$, (ii) the quasilinear phase $t_{sat}<t<\tau_{NL}$ and (iii) the strongly nonlinear phase $t>\tau_{NL}$.
The characteristic time scales $t_{sat}$ and $\tau_{NL}$ are highlighted in Fig.~\ref{fig:electron_acc} by vertical dashed lines. The latter time scale constitutes the core of the full-kinetic modelling, therefore, an extensive discussion is provided in Sec.~\ref{subs:modified_QL}.
In the linear phase, the electric field grows exponentially as predicted from linear theory. As expected, no electron acceleration is observed.
In the quasilinear phase, electron acceleration starts to take place as observed in both simulations. In this phase the efficiency of the acceleration is directly related to the driver intensity (i.e. the stronger the electric field the stronger the acceleration), as expected from the QL diffusion equations (Eqs.~\ref{eq:QL_cairns_fe}-\ref{eq:QL_cairns_De}).
Finally, in the strongly nonlinear phase, although the electric field amplitude remains constant and finite, the acceleration stops due to the onset of nonlinear LD-like effects.
Such a multi-phase evolution is observed in both strong and weak gradient simulations (see Fig.~\ref{fig:electron_acc}).
A main difference between the two simulations is the presence -- in the strong gradient simulation -- of an overshoot in the electric energy, not observed in the weak gradient simulation. 
Such overshoot corresponds to a sharp peak in the electric-to-thermal energy ratio at $t\approx t_{sat}$, which subsequently leads to a strong electron acceleration for the overshoot duration (see Fig.~\ref{fig:electron_acc}, left panel around $t_{sat}$). However, the overall contribution of this phenomenon to the total amount of accelerated electrons is shown not to be significant in Sec.~\ref{sec:discussion}.
In the end, for both simulations, the supra-thermal electron density in the inhomogeneous layer is increased by around $15\%-20\%$.
However, in the two cases this same final value is attained through a different evolution: in the strong gradient simulation the acceleration is faster but stops sooner (at $\tau_{NL}\approx10$), while in the weak gradient simulation the acceleration occurs at a slower rate but remains efficient on longer time scales (up to $\tau_{NL}\approx40$). Interestingly, these two effects seems to compensate each other.

The nonlinear time scale $\tau_{NL}$, not accessible to quasilinear models, represents the original result of our full-kinetic simulations; an extensive discussion is left to Sec.~\ref{sec:discussion}.

\subsection{Results of the standard quasilinear model}\label{subs:results_ql}

To enable a quantitative comparison between the QL model described in Sec.~\ref{subs:ql_model} and the full-kinetic one, the former is solved using the same plasma parameters as for the unstable layer of the full-kinetic simulations. More precisely, the input parameters of the QL model are obtained by averaging the plasma density, and the magnetic field in the layer $3.5<x<4$ (resp. $3.5<x<4.5$) for the strong (resp. weak) gradient full-kinetic simulation. 
    
\begin{figure}
    \centering
    \includegraphics[width=\linewidth]{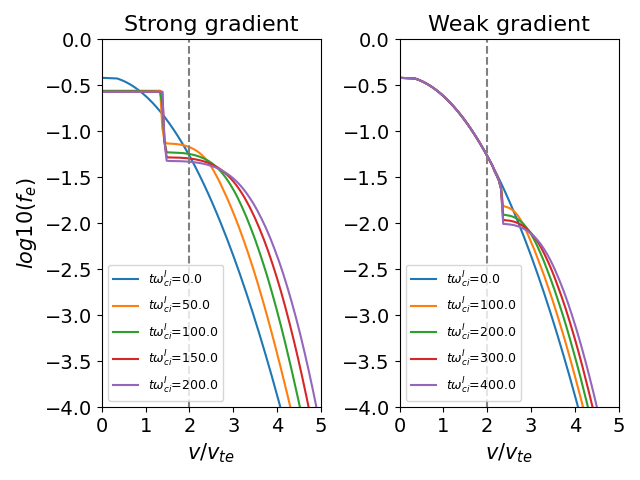}
    \caption{Results of the QL model for the strong and weak gradient parameter case, left and right panel, respectively. Both panels show the electron distribution function at different time instants in the direction parallel to the magnetic field. Grey vertical dashed lines at $v=2v_{the}\approx v_{phz}$ corresponds to the point from where the supra-thermal electron density (Eq.~\ref{eq:tracer}) is computed }
    \label{fig:QL_Solution_fe}
\end{figure}

Numerical integration of the QL model (Eqs.~\ref{eq:QL_cairns_fe}-\ref{eq:QL_cairns_gammae}) provides the time evolution of the electron distribution function $f_e(v_z)$ shown in Fig.~\ref{fig:QL_Solution_fe} at different time instants. We observe a diffusion in velocity space for both ``strong'' and ``weak'' gradient cases, corresponding to electron acceleration by LHW. The characteristic timescale for such acceleration turns out to be of the order of hundreds of ion cyclotron periods. As expected, this time scale is longer in the weak gradient case than in the strong gradient one.

Eventually, electron acceleration described by QL theory slows down after long times. Indeed, as already explained in~\citet{Cairns2005}, the characteristic time scale to accelerate an electron of velocity $v_\parallel$ by an amount $\delta v_\parallel$ scales as~$\sim v^5_\parallel$ for~$\delta v_\parallel {\sim} v_\parallel$. 
With the parameters used in our two simulations, the electron acceleration obtained from the QL model becomes negligible after time $\tau_{Diff} \gtrapprox 150$ (resp. $\tau_{Diff} \gtrapprox 400$) in the strong (resp. weak) gradient case, as shown in Fig.~\ref{fig:QL_Solution_fe}.
Such LHDI electron acceleration appears much weaker than those presented in the work of~\citet{Cairns2005}. 
This discrepancy is directly associated to the choice of the parameter $S_{max}/nT_i$, the electric-to-thermal energy at the saturation of the LHDI, which is proportional to the quasilinear diffusion coefficient (Eq.~\ref{eq:QL_cairns_De}). 
In this study, we use $S_{max}/nT_i = 0.01$ (resp. $0.002$) for the strong (resp. weak) gradient case, to obtain the results shown in Fig.~\ref{fig:QL_Solution_fe}. These values are obtained from Eq.~(\ref{eq:SnTi_saturation}) and are consistent with those observed in our full-kinetic simulations.
Differently, in the seminal paper of~\citet{Cairns2005} the authors have used a value $S_{max}/nT_i=0.5$ that is two orders of magnitude larger than what is expected from the theory of saturation of the LHDI in~\citet{Davidson1978} using the configuration considered in that study. 
Such choice of an overestimated value of the parameter $S_{max}/nT_i$ unavoidably leads to an overestimation of the quasilinear diffusion coefficient, and therefore of an higher electron acceleration efficiency.
This points out that a consistent choice of the electric-to-thermal energy ratio at saturation is crucial in quasilinear models in order to accurately assess a reliable value of the quasilinear diffusion coefficient. 

Despite the different choice of the parameter $S_{max}/nT_i$, the time evolution of the electron distribution function obtained with the QL model agrees qualitatively with the one presented in~\citet{Cairns2005}.

Nonetheless, the comparison with the evolution obtained from the full-kinetic model is left to the next section (Sec.~\ref{subs:comparison_ql_pic}).   

\section{Discussion: towards an extended quasilinear model and beyond}\label{sec:discussion}

In this section, we first compare the full-kinetic and standard quasilinear results, we show the discrepancies between the two models, and we highlight the physical processes that give rise to such discrepancies. 
Then, in order to overcome the intrinsic limits of the standard quasilinear theory, we build an extended quasilinear model (eQL) that includes the consequences of nonlinear LD-like effects. 
Finally, we validate such eQL and extrapolate how it scales with the plasma parameters of interest (e.g. ion-to-electron mass ratio, gradient length) in order to enable its use beyond the two specific set of parameters of our full-kinetic simulations.

\subsection{Comparison between the full-kinetic and the QL models}\label{subs:comparison_ql_pic}

\begin{figure}
    \centering
    \includegraphics[width=\linewidth]{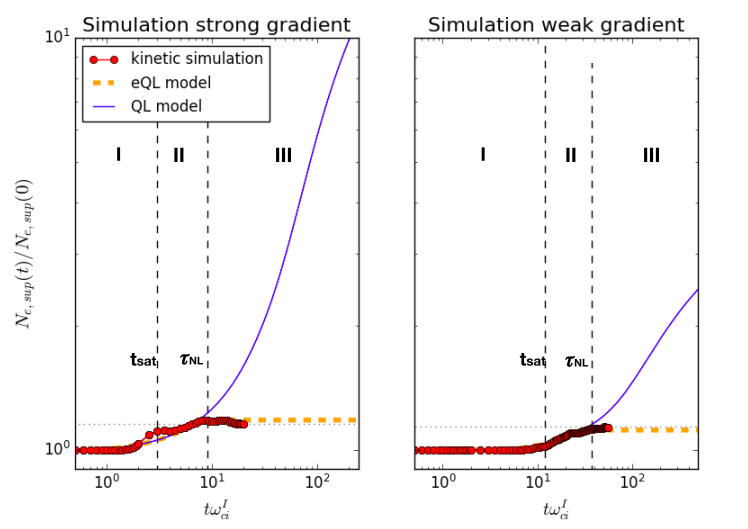}
    \caption{Comparison of the evolution of supra-thermal electron density (tracer of LHDI electron acceleration) computed from full-kinetic simulations (red curve), QL model (blue curve), extended QL model (orange curve). Vertical dashed lines indicate the saturation time $t_{sat}$, and nonlinear time $\tau_{NL}$. Horizontal dotted grey lines indicate the results of Eq.~(\ref{eq:Delta_Nesup_fin_chi}). Note that time axes are different for the two simulations.}
    \label{fig:QL_comparison}
\end{figure}

To quantitatively compare the results of the full-kinetic and QL models on LHDI electron acceleration, we focus on the evolution of the supra-thermal electron density (defined in Eq.~\ref{eq:tracer}), shown for both strong and weak gradient cases in Fig.~\ref{fig:QL_comparison}, where the two characteristic times $t_{sat}$ and $\tau_{NL}$ -- previously identified from the full-kinetic simulations -- are reminded for sake of clarity (vertical dashed lines). 
The three-phase evolution of LHDI electron acceleration observed in the full-kinetic simulations (red lines in Fig.~\ref{fig:QL_comparison}) is not explicable in terms of the QL model (blue lines in Fig.~\ref{fig:QL_comparison}). 
Two main discrepancies are identified.

First, we observe a minor discrepancy between the two models around $t = t_{sat}$ in the strong gradient case, see Fig.~\ref{fig:QL_comparison} left panel. At this time, the supra-thermal electron density obtained from the full-kinetic simulation (red curve) is higher than that obtained from the QL model (blue curve). 
This enhanced electron acceleration is due to the overshoot of the electric field in the full-kinetic simulation, effect not included in the QL model, and not observed in the weak gradient case. However, this short and sharply peaked phenomenon brings a negligible contribution to the total amount of supra-thermal electrons at the end of the simulation. 

Second, we observe a strong discrepancy between the two models for times $t>\tau_{NL}$ (phase III in Fig.~\ref{fig:QL_comparison}).
Indeed, on the one hand, the electron acceleration stops at time $t\approx \tau_{NL}$ in full-kinetic simulations (red curve) due to the onset of nonlinear LD-like effects, while on the other hand, the electron acceleration goes on in the QL model (blue curve) up to time $t\approx \tau_{Diff}$.
The fact that the value of $\tau_{NL}$ (obtained from full-kinetic simulations) is about one order of magnitude lower than $\tau_{Diff}$ (obtained from QL simulations) is the main reason for the strong discrepancy between the two models.
Such discrepancy is due to the fact that the QL model does not include the nonlinear LD-like effects that are responsible for the abrupt stop of electron acceleration observed at $t=\tau_{NL}$ in the full-kinetic simulations.
Thus, even though the two models are in good agreement in the linear and quasilinear phases (phases I and II in Fig.~\ref{fig:QL_comparison}), the final value of the supra-thermal electron density is strongly overestimated by the QL model.
In the following, we shall use the increase in the supra-thermal electron density, defined as:
\begin{equation}\label{eq:def_Delta_Nesup}
    \Delta N_{e,sup}=N_{e,sup}(\tau_{NL})-N_{e,sup}(0)
\end{equation}
to quantify the discrepancy between the models. In particular, the value of $\Delta N_{e,sup}$ is 50 (resp. 20) times higher in the strong (resp. weak) gradient case for the QL model as compared to the full-kinetic model.

\subsection{Need for an extended quasilinear model}\label{subs:modified_QL}

Quasilinear theory is a powerful tool to study wave-particle interaction in an analytical framework due to its relative simplicity. Therefore, it represents an ideal way to provide quantitative predictions on LHDI electron acceleration/heating. 

However, due to its derivation via a perturbative approach, QL theory does not include ``strong'' nonlinear effects (i.e. those effects, arising at sufficiently long times, that alter the wave fluctuations due to the feedback from the modified distribution function to the fields).
Although they are not included in the QL model, these effects are well reproduced by the full-kinetic simulations given its \textit{ab initio} nature. 
The comparison between both models (Sec.~\ref{subs:comparison_ql_pic}) highlights the need to build an extended QL model (hereafter called eQL) that overcomes the intrinsic limitations of standard QL model and provide a dynamics consistent with that observed in full-kinetic models. 
In this purpose, we argue that such eQL model:
\begin{itemize}
    \item does not require to include the possible overshoot of the electric field, as we have shown it is not a dominant process in energizing electrons, 
    \item requires to include the eventual inhibition of LHDI electron acceleration by nonlinear LD-like effects through the parameter $\tau_{NL}$.
\end{itemize}

\begin{figure}
    \centering
    \includegraphics[width=\linewidth]{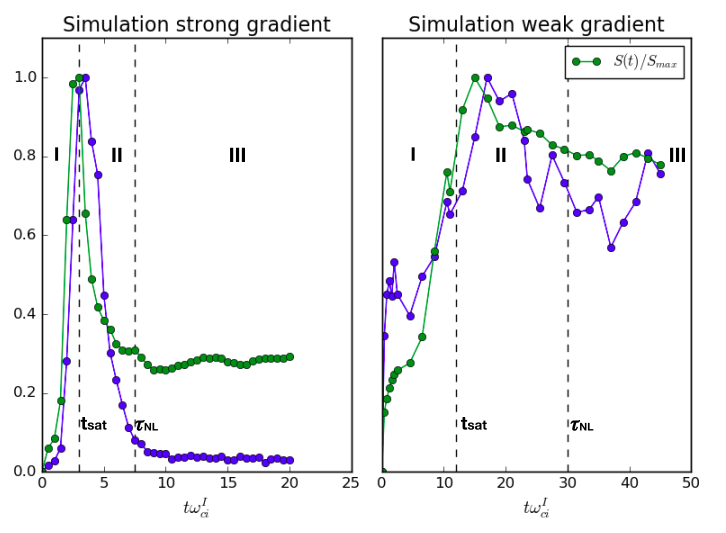}
    \caption{Evolution of the electric field energy $S(t)/S_{max}$ (green curve), and of the QL diffusion coefficient $D_e(t, v_z=2v_{the})/D_{e,max}$ in Eq.~(\ref{eq:QL_cairns_De}) (blue curve), computed for both full-kinetic simulations. Note that time axes are different for the two simulations.}
    \label{fig:NL_diffusion}
\end{figure}

The nonlinear time $\tau_{NL}$ is the parameter used in the eQL to define when nonlinear LD-like effects become dominant in the layer, inhibiting efficient wave-particle interactions.
From such time, the QL diffusion coefficient (Eq.~\ref{eq:QL_cairns_De}) becomes negligible even though the amplitude of the electric field remains constant, as shown in Fig.~\ref{fig:NL_diffusion} for both full-kinetic simulations. This is due to its integral dependence on $k^2_{\parallel}/k^2_{\bot}$. 
The underlying physical process is the energy transfer, in $k$-space, from resonant, oblique modes to out-of-resonance, more perpendicular modes.
Therefore, we explicitly switch off the QL diffusion coefficient for $t>\tau_{NL}$ in the eQL model (Eq.~\ref{eq:eQL_fe}-\ref{eq:eQL_De}).

We now focus on the estimation of this new parameter $\tau_{NL}$. 
Previous works on the LHDI, addressing its nonlinear evolution using 2D full-kinetic simulations, suggest a connection between LHDI modes and drift-kink-instability (hereafter DKI) modes~\citep{Pritchett1996,Shinohara1999,Lapenta2002}.
DKI modes are characterized by (i) frequency and growth rate $\gamma_{_{DKI}}$ smaller than the ion gyrofrequency and (ii) wavevectors perpendicular to the magnetic field~\citep{Daughton1998,Daughton1999}.
Following these studies, we assume that the mechanism underlying the energy transfer from oblique to strictly perpendicular modes at $t>\tau_{NL}$, observed in our full-kinetic simulations, is a LHDI-DKI coupling.
Consequently, we estimate that the time of onset of LHDI nonlinear LD-like effects scales as the inverse linear growth rate of the DKI, i.e. $\tau_{NL} \propto 1/\gamma_{_{DKI}}$.
The DKI linear growth rate $\gamma_{_{DKI}}$ scaling with plasma physical parameters has been addressed in~\citet{Daughton1998} using kinetic theory, and validated in~\citet{Daughton1999} using 2-fluid theory. In our work, we use such results to build our eQL model (Eq.~\ref{eq:eQL_tauNL}).

Taken all those considerations into account, the eQL model reads:
\begin{eqnarray}
    \partial_t f_e(v_{\parallel},t) & = &\label{eq:eQL_fe}
                \partial_{v_{\parallel}} \mathcal{D}_{NL} \partial_{v_{\parallel}} f_e\\
               \mathcal{D}_{NL}(v_{\parallel},t) & = &\label{eq:eQL_De}
                \begin{cases}
                D_e(v_{\parallel},t) &t<\tau_{NL}\\
                0                    &t \geq \tau_{NL}
                \end{cases}\\
              \tau_{NL}            & = &\label{eq:eQL_tauNL}
                            \tau_0 \left( \frac{m_i}{m_e} \right)^{1/2} \left( \epsilon_n\rho_i \right)^{-2} \left( 1 + \frac{T_e}{T_i}\right)^{-1}
\end{eqnarray}
where $\tau_0=1.5$ is a constant obtained from our two full-kinetic simulations. This value is obtained by fitting the evolution of the supra-thermal electron density of the full-kinetic simulations with the one of the eQL model.

\subsection{Validation of the extended quasilinear model}\label{subs:validation_modified_QL}

The eQL presented in Sec.~\ref{subs:modified_QL} (Eqs.~\ref{eq:eQL_fe}-\ref{eq:eQL_tauNL}) retains all the advantages of the standard QL model (namely being analytical and easily integrable using a numerical solver) while extending its range of validity to asymptotically long time scales. This now enables to provide quantitative predictions of LHDI electron acceleration. Indeed, as shown by the time evolution of the supra-thermal electron density in Fig.~\ref{fig:QL_comparison}, unlike the standard QL predictions (blue), the eQL model (orange) is able to reproduce the predictions of the nonlinear Vlasov-Maxwell theory (red). 
The relative discrepancy between the two curves does never exceed 5\% in both simulations.
This validation of the eQL model is of particular interest to quantify the LHDI electron acceleration over long times.

The results of the eQL model strongly depend on the estimation of the nonlinear time $\tau_{NL}$, identified as the inverse linear growth rate of the DKI $1/\gamma_{_{DKI}}$. This choice is justified both (i) \textit{a priori}, by previous numerical works that showed evidence that the long time evolution of a LHDI-unstable layer gets coupled to a DKI~\citep{Pritchett1996,Shinohara1999}; and (ii) \textit{a posteriori}, by validating that the scaling $\gamma_{_{DKI}}\propto(\epsilon_n\rho_i)^2$ in Eq.~(\ref{eq:eQL_tauNL}) is consistent with the outputs of our two full-kinetic simulations.

In our two full-kinetic simulations -- that have enabled to both define and validate the eQL model -- we have used a reduced ion-to-electron mass ratio $m_i/m_e=100$.
Using such a reduced mass ratio is a standard procedure in PIC plasma simulations, since it allows to reduce the scale separation among the two species in order to run the numerical simulation on reasonable amount of CPU time (still about one million computational hours per simulation, in our case), while maintaining the necessary temporal and spatial scales separations between both species dynamics. 
However, the properties of the LHDI actually depend on the ion-to-electron mass ratio, because of the hybrid character of the instability. Therefore, the quantitative predictions of our full-kinetic simulations are not directly applicable to a realistic plasma configuration (with a physical proton-to-electron mass ratio $1836$). This is exactly where an eQL model becomes extremely useful.

\subsection{Using the extended quasilinear model to assess LHDI electron acceleration at physical mass ratio} \label{subs:use_modified_QL}

\begin{figure*}
    \centering
    \includegraphics[width=0.9\linewidth]{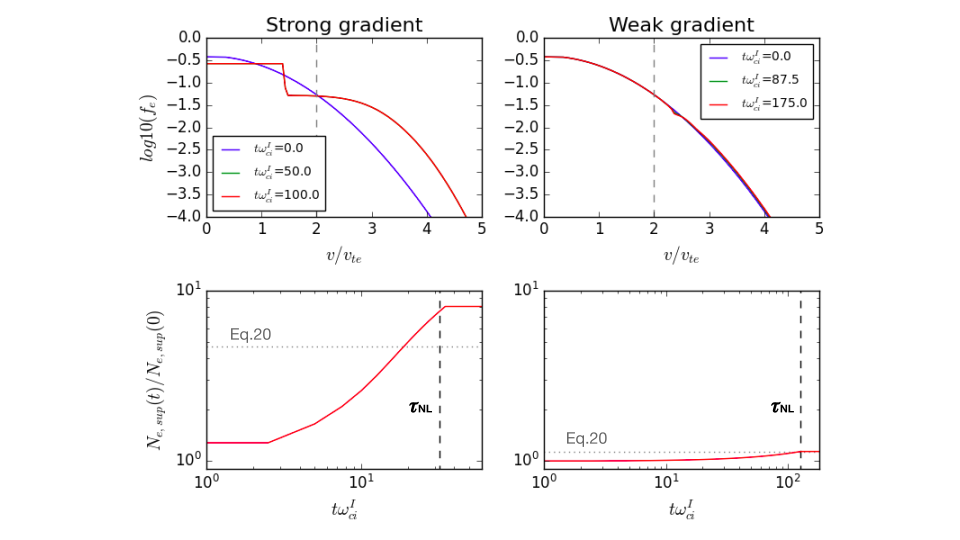}
    \caption{Output of the eQL model using the strong (left panels) and weak (right panels) set of plasma parameters with realistic proton-to-electron mass ratio.
    Top panels: same as Fig.~\ref{fig:QL_Solution_fe}.
    Bottom panels: evolution supra-thermal electron density, see Eq.~(\ref{eq:tracer}).
    Vertical dashed line indicate the nonlinear time $\tau_{NL}$ in Eq.~(\ref{eq:eQL_tauNL}). Horizontal dotted grey lines indicate the results of Eq.~(\ref{eq:Delta_Nesup_fin_chi}). Note that time axes are different for the two simulations.}
    \label{fig:eQL_mime_real}
\end{figure*}

The eQL model does not suffer from the strong computational constraints of full-PIC simulations and enables us to address, in this section, the question of LHDI electron acceleration using a realistic proton-to-electron mass ratio.
The results of the eQL models are summarized in Fig.~\ref{fig:eQL_mime_real}, showing both the electron distribution function evolution averaged in the inhomogeneous layer (top panels) and the resulting supra-thermal electron density (bottom panels), for both strong and weak gradient setups. Here, we use the two setups described in Sec.~\ref{sec:models}, where only the mass ratio is modified to address the case $m_i/m_e = 1836$. 
We emphasize that full kinetic simulations using such a physical mass ratio would have been extremely challenging computationally, so that we consider the eQL model as a way to extrapolate the results of our full-kinetic simulations (done using a reduced mass-ratio) to a physical plasma environment.

Compared to results shown in Sec.~\ref{subs:validation_modified_QL} for a reduced mass ratio, using a physical mass ratio the nonlinear time increases following Eq.~(\ref{eq:eQL_tauNL}). This means that the LHW-particle interaction occurs over longer times, thus more electrons are accelerated.
However, on the one hand, this effect can be compensated by the reduction of the electric field energy of order $\sim m_e/m_i$ (Eq.~\ref{eq:SnTi_saturation}) if the LHDI saturates through current relaxation, i.e. in the weak gradient case, see Fig.~\ref{fig:eQL_mime_real} right panel.
On the other hand, if the LHDI saturates through ion trapping the saturation level does not depend on the mass-ratio (Eq.~\ref{eq:SnTi_saturation}), therefore, the LHDI electron acceleration is more efficient by a factor $\sim m_i/m_e$ due to the increase in the nonlinear time.
All in all, the mass ratio does not affect the weak gradient case, while it increases the fraction of LHDI accelerated electrons by a factor $\sim 1836/100$ in the strong gradient case.
These predictions -- solely based on scaling arguments -- are well reproduced by the numerical solution of the eQL using a realistic proton-to-electron mass-ratio, as shown in Fig.~\ref{fig:eQL_mime_real}, and are also confirmed by the analytical estimates developed below in this section.

Now, we focus on the value of the supra-thermal electron density at asymptotically long times, since, all in all, this is the quantity relevant for space plasma observations of LHDI accelerated electrons. 
Under some simplifying assumptions we derive an analytical expression for this quantity in the framework of the eQL model.

First, we approximate the evolution of the supra-thermal electron density by a linear interpolation in the time interval $t_{sat}<t<\tau_{NL}$ to get
\begin{equation}
\label{eq:Delta_Nesup}
    \Delta N_{e,sup} = N_{e,sup}(\tau_{NL}) - N_{e,sup}(0) = (\tau_{NL}-t_{sat}) \frac{d}{dt} N_{e,sup}(t)
\end{equation}
then, we integrate Eq.~(\ref{eq:QL_cairns_fe}) in $v_\parallel$-space to get
\begin{equation}
\label{eq:dNsup_dt}
    \frac{d}{dt} N_{e,sup}(t) = \left[ D_e \partial_v f_e \right]_{v_{\parallel}=2v_{the}}
\end{equation}
which is constant in the interval $t_{sat}<t<\tau_{NL}$ since we assume that (i) the distribution function is weakly modified by the interaction with the wave, i.e. $f_e(t,v_{\parallel}) \approx f_e(0,v_{\parallel})$, and (ii) the amplitude of the resonant electric field wave $|E(k_\bot,k_\parallel)|$ remains constant after the saturation of the LHDI for $t>t_{sat}$.
Finally, under the previous assumptions, using the expression for the QL diffusion coefficient (Eq.~\ref{eq:QL_cairns_De}), assuming a Maxwellian distribution function for the electrons, the Eqs.~(\ref{eq:Delta_Nesup}-\ref{eq:dNsup_dt}) lead to
\begin{equation}
\label{eq:Delta_Nesup_fin}
    \frac{\Delta N_{e,sup}}{N^0_{e,sup}} = 0.1\left( \frac{\tau_{NL}-t_{sat}}{\omega^{-1}_{ci}}\right) \left( \frac{\omega^2_{pe}}{\omega^2_{ce}} \frac{S_{max}}{n_0T_i} \sqrt{\frac{m_i}{m_e}} \left( \frac{T_i}{T_e}\right)^{3/2}
    \right)
\end{equation}
Before going on, we stress here that the increase in supra-thermal density $\Delta N_{e,sup}$ is proportional to the amplitude of the electric field at saturation $S_{max}$. This emphasizes how the input parameter $S_{max}/nT_i$ impacts the output of QL theory and further supports the discussion regarding the difference between our results and the ones by~\citet{Cairns2005} in Sec.~\ref{subs:results_ql}.  

Finally, expressing the electric field at saturation $S_{max}$ using Eq.~(\ref{eq:SnTi_saturation}), the nonlinear time $\tau_{NL}$ using Eq.~(\ref{eq:eQL_tauNL}), the saturation time $t_{sat}$ using Eq.~(\ref{eq:saturation_time}), and the LHDI growth rate $\gamma_{_{LHDI}}$ using Eq.~(\ref{eq:gammaL_saturation}), and under the assumption $\omega_{pe}>\omega_{ce}$, the supra-thermal density increase in Eq.~(\ref{eq:Delta_Nesup_fin}) becomes:
\begin{equation}
\label{eq:Delta_Nesup_fin_chi}
    \frac{\Delta N_{e,sup}}{N^0_{e,sup}}  =
        \begin{cases}
        0.3 \left( 1 + \frac{T_e}{T_i} \right)^{-1} (1 - \mathcal{\chi}\frac{m_e}{m_i}) 
        ~~~~~~~~~~~\text{current relaxation}\\
        
        0.004 \left( 1 + \frac{T_e}{T_i} \right)^{-1} (1 - \mathcal{\chi}\frac{m_e}{m_i}) \frac{m_i}{m_e} (\epsilon_n \rho_i)^3 ~~\text{ion trapping}
         \end{cases}
\end{equation}
where $\chi$ reads:
\begin{equation}
  \label{eq:Nesup_chi}
  \mathcal{\chi} = \frac{1.6}{\tau_0} \left(1+\frac{T_e}{T_i}\right)\sqrt{1+\beta_i/2} \ln\left[S_{max}/S(t=0)\right] 
\end{equation}
With the parameters considered in this study, and typically encountered in space plasmas, $\mathcal{\chi} \frac{m_e}{m_i} \ll 1$. As a consequence, the supra-thermal electron density increase at long times is well approximated analytically by
\begin{equation}
\label{eq:Delta_Nesup_fin_fin}
    \frac{\Delta N_{e,sup}}{N^0_{e,sup}} =
        \begin{cases}
        0.3 \left( 1 + \frac{T_e}{T_i} \right)^{-1} ~~~~~~~~~~~ \text{current relaxation}\\
        
        0.004 \left( 1 + \frac{T_e}{T_i} \right)^{-1} \frac{m_i}{m_e} (\epsilon_n \rho_i)^3  ~~ \text{ion trapping}
         \end{cases}
\end{equation}
depending on the LHDI saturation mechanism at play. 

This estimation leads to a total increase in the supra-thermal electron density summarized in Tab.~\ref{tab:table_sec4} for the different sets of parameters used throughout this work.
These values are also shown as horizontal grey dotted lines in both panels of Fig.~\ref{fig:QL_comparison} -- for reduced mass ratio -- and Fig.~\ref{fig:eQL_mime_real} for a physical mass ratio.
\begin{table}
    \caption{Values of $\Delta N_{e,sup}/N^0_{e,sup}$ for the different plasma configurations considered in this study, computed using the eQL model outlined in Sec.~\ref{subs:modified_QL} and the approximated analytical expression in Eq.~(\ref{eq:Delta_Nesup_fin_chi}). The last column shows the relative discrepancy between these two results.}
    \label{tab:table_sec4}
    \centering
    \begin{tabular}{| c c | c c c |}
        \hline 
        gradient $\epsilon_n \rho_i$  & $m_i/m_e$  &  eQL model  &  Eq.~(\ref{eq:Delta_Nesup_fin_chi})  &  error [\%] \\
        \hline
        1   &  100   & 0.18      & 0.16 & 12\\
        0.5 &  100   & 0.12      & 0.14 & 15\\
        \hline
        1   &  1836  & 7.62      & 3.67 & 70\\
        0.5   &  1836  & 0.14      & 0.15 & 7\\
        \hline
    \end{tabular}
\end{table}
From the values in Tab.~\ref{tab:table_sec4}, we infer that our analytical approximation (Eq.~\ref{eq:Delta_Nesup_fin_chi}) is valid (error of the order of $10\%$) in the limit of ``weak'' LHDI electron acceleration (i.e. $\Delta N_{e,sup}/N^0_{e,sup} \lessapprox 1$).
Stronger discrepancies arise for stronger electron acceleration (see third row in Tab.~\ref{tab:table_sec4}), as expected.

\subsection{Application to Mercury's magnetopause}

Mercury's magnetopause represents an excellent ``textbook'' example of a plasma boundary with a ion kinetic scale density gradient potentially LHDI-unstable. 
Previous space missions at Mercury -- Mariner 10~\citep{Russell1988} and MESSENGER~\citep{Solomon2007} -- did not bring an instrumental payload capable of providing simultaneous measurements of electric field in the lower-hybrid frequency range and of the electron distribution function. Therefore, the expected LHW physics at Mercury cannot yet be tackled from past observations.  
However, the physics at these scales/range is exactly one of the main scientific objectives of the ongoing ESA/JAXA space mission BepiColombo~\citep{Benkhoff2010}. 
Different plasma instruments onboard the Mio spacecraft will provide measurements of the electron distribution function (MPPE/MEA, Mercury Plasma Particle Experiment / Mercury Electron Analyzer) and of the electric field in the lower-hybrid frequency range (PWI, Plasma Wave Investigation) enabling the possibility to directly address the physics of Mercury's strongly inhomogeneous magnetopause. 
In this section, we quantify the expected efficiency of LHDI electron acceleration at Mercury's magnetopause using the eQL model developed in our work in support for future BepiColombo observations.

As previously pointed out in Sec.~\ref{subs:fullkinetic_model}, the parameters used in our strong and weak gradient cases are taken in the range expected at Mercury's magnetopause~\citep{Slavin2008,Gershman2015}.
Therefore, we make use of the results of our eQL model with physical mass ratio in Sec.~\ref{subs:use_modified_QL} to assess the importance of electron acceleration driven by the LHDI at Mercury's magnetopause.

First, we discuss the features of the LHW that are expected to be generated by the development of the LHDI at Mercury's magnetopause.
These waves have frequency $f\approx f_{_{LH}} \approx $ 50-100 Hz and wavelength $\lambda \approx 2\pi / \rho_e \approx$ 5-15 km (using hereafter typical magnetic field values at Mercury's magnetopause of 10-30 nT, ion temperature 30-70 eV, and density 10-30 cm$^{-3}$, see~\citet{Milillo2020}). Assuming that those waves are advected by the shocked solar wind flow with a speed $V_{_{SW}} \approx$ 100 km/s, the Doppler-shifted frequency of the LHW lies in the range $f' = f \pm k V_{_{SW}} \approx$ 0-200 Hz. 
Moreover, the electric field amplitude of these LHW at saturation is $E \approx$ 10-100 mV/m, as obtained from Eq.~(\ref{eq:SnTi_saturation}).
These typical frequency and energy range of LHW at Mercury suggests the use of electric field instruments onboard Mio spacecraft to address the LHW physics at Mercury's magnetopause. In particular, the sensors MEFISTO and WPT -- part of the PWI consortium -- can provide electric field observations in this frequency range with a sensitivity of the order of $10^{-3}$ mV/m, well-below the expected amplitude of these waves~\citep{kasaba2020}. 
\begin{figure*}
    \centering
    \includegraphics[width=0.7\linewidth]{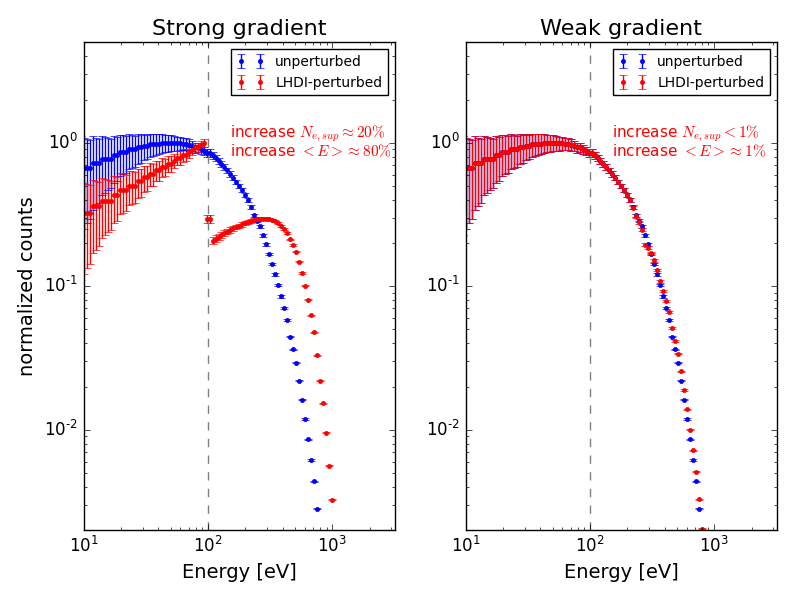}
    \caption{Instrumental response of the instrument MPPE/MEA simulated using our extended QL model for both strong $\epsilon_n \rho_i{=}1$ and weak $\epsilon_n \rho_i{=}0.5$ gradient cases (left and right panels), in the direction parallel to the magnetic field.
    Unperturbed electron distribution function using temperature $T_e=$50 eV in blue, and results of eQL model at time $t=\tau_{NL}$ in red. The dashed vertical line indicates the energy from which the supra-thermal electron density $N_{e,sup}$ is computed.  
    }
    \label{fig:spectra_MEA}
\end{figure*}

Second, we discuss the features of electron distribution functions possibly modified by resonant interaction with the previously discussed LHW.
This resonant wave-particle interaction accelerates sub-thermal electrons (with speed $v_z \lessapprox v_{the}$) to supra-thermal energies (with speed $v_z \gtrapprox 2v_{the}$) in the direction parallel to the ambient magnetic field, as shown in Fig.~\ref{fig:eQL_mime_real} top panels.
In the following we assume an unperturbed electron temperature of 50 eV (typical values at Mercury's magnetopause being $\sim20-100$ eV~\citep{Ogilvie1974}).
This acceleration process is well in the range of observations of the electron instrument MPPE/MEA onboard the Mio spacecraft of the BepiColombo mission~\citep{Saito2010}, since this instrument includes two electron analyzers that can measure the three-dimensional energy distribution of electrons in the range 3-3000 eV (in solar wind mode) or 3-25500 eV (in magnetospheric mode), with a time resolution of 1 second~\citep{Milillo2020}.
Here, we simulate the instrumental response of MPPE/MEA when encountering electrons accelerated by a LHDI at Mercury's magnetopause, as shown in Fig.~\ref{fig:spectra_MEA}, with the unperturbed electron distribution function in blue, and the one resulting from interaction with LHDI at long times in red. 
In Fig.~\ref{fig:spectra_MEA}, the uncertainties on the simulated response of the sensor MEA are obtained using the uncertainty on the G-factor of the instrument reported in~\citet{Saito2010} (of the order of 10\%).
So, on the one hand we predict that the modifications in the electron distribution function above 100 eV are observable by MPPE/MEA in the case of a strongly inhomogeneous magnetopause (width around one ion gyro-radius, or less), as shown in the left panel of Fig.~\ref{fig:spectra_MEA}.
On the other hand, since this process is very sensitive to the width of the magnetopause layer, we expect negligible modifications in the electron distribution function for less inhomogeneous magnetopause conditions (width around two ion gyro-radii or more), as shown in the right panel of Fig.~\ref{fig:spectra_MEA}.
With typical magnetic field and temperature values at Mercury's magnetopause the ion gyroradius is around $20-80$ km.

We also assess that the interaction between LHDI and electrons increases both (i) the density of supra-thermal electrons $N_{e,sup}$ (i.e. electrons with energy higher than 100 eV, see Eq.~\ref{eq:tracer}), and (ii) the electron temperature. In the simulated responses with strong (resp. weak) density gradient, the former increases by 20\% (resp. 0.3\%), due to the interaction with the LHW, and the latter increases by 80\% (resp. 1.5\%).
We stress here how these two scalar quantities could be used as tracers of LHDI-electron interaction events in MPPE/MEA data, and how this response is limited to the direction parallel/anti-parallel to the ambient magnetic field (i.e. around $\theta=0^{\circ}$ and $180^{\circ}$ in the electron pitch-angle distributions).

\begin{figure*}
    \centering
    \includegraphics[width=0.8\linewidth]{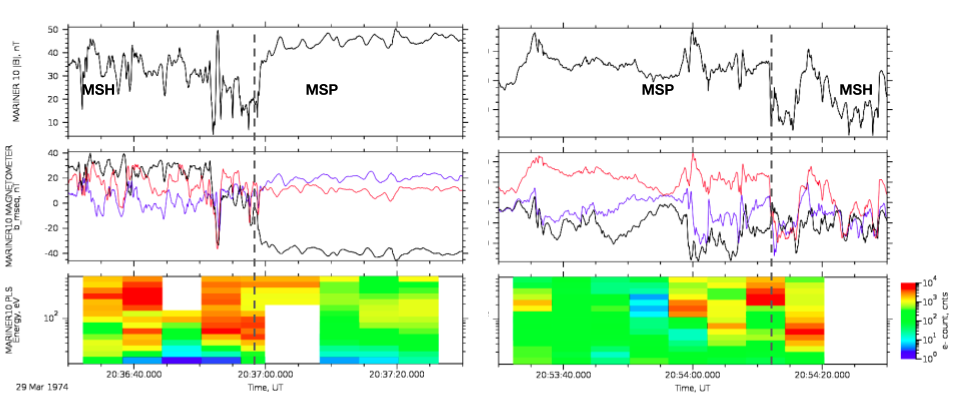}
        \includegraphics[width=0.7\linewidth]{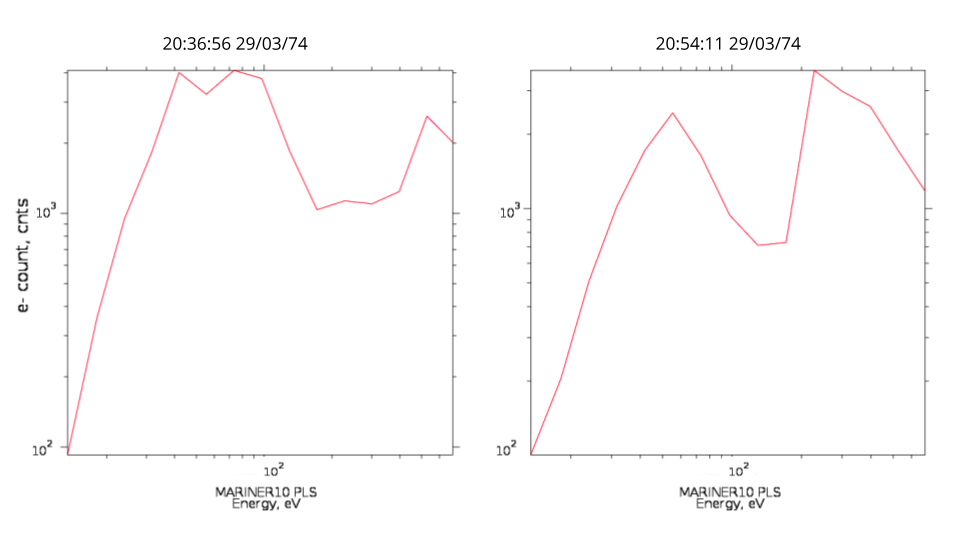}
    \caption{Mariner 10 observations during its first Mercury flyby on the 29 March 1974, inbound (left panels) and outbound (right panels) magnetopause crossings are shown, the associated time intervals are 20:36:30-20:37:30 and 20:53:30-20:54:30 respectively. Magnetosphere (MSP) and magnetosheat (MSH) plasma measurements are shown. From top to bottom, we show the magnetic field module (first row), the magnetic field components (second row), the electron energy spectra as a function of time (third rows), and the electron energy spectrum at the time of the crossings (fourth row) -- time corresponding to the grey vertical dashed lines in the first three panels.}
    \label{fig:Mariner10}
\end{figure*}

Third, we present electron observations made by the NASA spacecraft Mariner 10 during its first Mercury flyby on 29 March 1974~\citep{Ogilvie1974}. In Fig.~\ref{fig:Mariner10} we show magnetic field (upper panels) and electrons (bottom panels) data for the inbound (left) and outbound (right) magnetopause crossings. In particular, we observe a bimodal character of the electron energy distribution during the crossing (bottom panels), with one ambient electron population around 50-70 eV and a second energized population around 300-500 eV. This is consistent with the 
signatures expected for LHDI-accelerated electrons.
Although these observations 
suggest that the LHDI might play a role in the electron energization at Mercury's Magnetopause, 
the Mariner 10 data cannot unambiguously confirm this hypothesis because of the lack of electric field observations. Moreover, the low time resolution of Mariner 10 electron measurements (1/6 $s^{-1}$), the narrow energy range (13.4-687 eV) and the lack of a statistically significant number of crossings make the Mariner 10 observations prevent from a conclusive evidence of LHDI-accelerated electrons at Mercury's magnetopause.

The limits of such Mariner 10 measurements will be overcame by the more advanced and complete payload 
of the ESA-JAXA BepiColombo space mission, especially the joint electric field observations of LHW (with PWI/MEFISTO and PWI/WPT) and electrons (with MPPE/MEA). We are therefore confident that the Mio plasma and electric field instrumental suites of BepiColombo will soon enable to shade light on the statistical relevance of LHDI and associated electron heating in the global dynamics of Mercury's magnetosphere frontier with the solar wind.

\section{Conclusions}\label{sec:conclusions}

In this work, we have addressed the question of electron acceleration efficiency by lower hybrid waves generated by the lower hybrid drift instability. 

For this purpose, we have performed 3D, full-kinetic numerical simulations to provide numerical evidence of electrons acceleration parallel to the ambient magnetic field by resonant wave-particle interaction with LHDI waves.
Our self-consistent nonlinear model has also enabled us to address the consequences of the saturation of the instability on the eventual stoppage of the electron acceleration. To our best knowledge, this is the first time that this process is self-consistently observed in full-kinetic simulations.
This represents the first original contribution of this work. 

Moreover, we have provided quantitative estimates of the efficiency of this resonant acceleration process. To model this process we have (i) used a standard QL model based on the work of~\citet{Cairns2005}, and (ii) developed an extended QL model that includes the consequences of nonlinear LD-like effects on long time scales evolution of the electron distribution function.
We have compared the results of these two quasilinear models and the full-kinetic model. 
Such comparison highlighted the limitations at long time scales of the standard quasilinear theory that paved the way for an extended one, designed to overcome such limitations.
Such comparison also enabled to validate the eQL model. 
This new extended quasilinear (eQL) model successfully captures the electron acceleration properties found on full-kinetic simulation results, and represents the second original contribution of this work. 

The eQL model, thanks to its simplicity, has enabled us to explore a range of parameter space that is not accessible to full-kinetic simulations due to computational constraints. In particular, we have addressed the efficiency of LHDI electron acceleration at Mercury's magnetopause, using a realistic proton-to-electron mass ratio.
In this context, we estimate that LHDI electron acceleration is an efficient mechanism to energize electrons during periods of strong magnetospheric compressions (when the magnetopause boundary steepens on scales of the order or lower than the ion gyroradius). This efficiency strongly depends on the density gradient at the magnetopause.
Under such conditions of ``steep'' magnetopause at Mercury, we expect strong signatures of LHDI electron acceleration in the BepiColombo instrumental suite response (PWI and MPPE/MEA).  

We are confident that our extended quasilinear model will enable to quantitatively study the efficiency of electron acceleration in inhomogeneous space plasmas, in support to past and future space plasma observations. 
More important, this work also provides a validated framework to extend the range of validity and applicability of quasilinear modeling, not only associated to lower hybrid drift instability -- as in this study -- but also to a broader variety of waves and instabilities in space plasmas.

\begin{acknowledgements}
This study was granted access to the HPC resources of SIGAMM infrastructure (http://crimson.oca.eu), hosted by Observatoire de la C\^{o}te d’Azur and which is supported by the Provence-Alpes C\^{o}te d’Azur region.
This work was performed using HPC resources from GENCI-CINES (Grant 2020-A0080411496).
We acknowledge PRACE for awarding us access to Joliot-Curie at GENCI@CEA, France.
We acknowledge the CINECA award under the ISCRA initiative, for the availability of high performance computing resources and support for the project IsC71.
We acknowledge the computing centre  of Cineca and INAF, under the coordination of the "Accordo Quadro MoU per lo svolgimento di attività congiunta di ricerca Nuove frontiere in Astrofisica: HPC e Data Exploration di nuova generazione", for the availability of computing resources and support.
The authors would like to thank the Smilei team for their brilliant technical support (https://smileipic.github.io/Smilei), and in particular Mickael Grech for the fruitful discussions.
Data analysis was performed with the AMDA science analysis system provided by the Centre de Données de la Physique des Plasmas (CDPP) supported by CNRS, CNES, Observatoire de Paris and Université Paul Sabatier, Toulouse. 
The Mariner 10 data is also available on the Planetary Data System (https://pds.nasa.gov).
We acknowledge the support of CNES for the BepiColombo mission.
\end{acknowledgements}

\bibliography{biblio}

\begin{thebibliography}{77}
\expandafter\ifx\csname natexlab\endcsname\relax\def\natexlab#1{#1}\fi

\bibitem[{Alexandrov {et~al.}(1984)Alexandrov, Bogdankevich, \&
  Rukhadze}]{Alexandrov1984}
Alexandrov, A.~F., Bogdankevich, L.~S., \& Rukhadze, A.~A. 1984, {Principles of
  Plasma Electrodynamics}, {Series in Electrophysics} ({Springer})

\bibitem[{{Alpers}(1969)}]{Alpers1969}
{Alpers}, W. 1969, Astrophysics and Space Science, 5, 425

\bibitem[{{Amatucci}(1999)}]{Amatucci1999}
{Amatucci}, W.~E. 1999, \jgr, 104, 14481

\bibitem[{{Andr{\'e}} {et~al.}(2001){Andr{\'e}}, {Behlke}, {Wahlund},
  {Vaivads}, {Eriksson}, {Tjulin}, {Carozzi}, {Cully}, {Gustafsson},
  {Sundkvist}, {Khotyaintsev}, {Cornilleau-Wehrlin}, {Rezeau}, {Maksimovic},
  {Lucek}, {Balogh}, {Dunlop}, {Lindqvist}, {Mozer}, {Pedersen}, \&
  {Fazakerley}}]{Andre2001}
{Andr{\'e}}, M., {Behlke}, R., {Wahlund}, J.~E., {et~al.} 2001, Annales
  Geophysicae, 19, 1471

\bibitem[{Andr{\'e} {et~al.}(2017)Andr{\'e}, Odelstad, Graham, Eriksson,
  Karlsson, Stenberg~Wieser, Vigren, Norgren, Johansson, Henri, Rubin, \&
  Richter}]{Andre2017}
Andr{\'e}, M., Odelstad, E., Graham, D., {et~al.} 2017, Monthly Notices of the
  Royal Astronomical Society, 469

\bibitem[{Bale {et~al.}(2002)Bale, Mozer, \& Phan}]{Bale2002}
Bale, S.~D., Mozer, F.~S., \& Phan, T. 2002, Geophysical Research Letters, 29,
  33

\bibitem[{{Benkhoff} {et~al.}(2010){Benkhoff}, {van Casteren}, {Hayakawa},
  {Fujimoto}, {Laakso}, {Novara}, {Ferri}, {Middleton}, \&
  {Ziethe}}]{Benkhoff2010}
{Benkhoff}, J., {van Casteren}, J., {Hayakawa}, H., {et~al.} 2010, Planetary
  Space Science, 58, 2

\bibitem[{Bernstein \& Engelmann(1966)}]{Bernstein1966}
Bernstein, I.~B. \& Engelmann, F. 1966, The Physics of Fluids, 9, 937

\bibitem[{Bingham {et~al.}(2002)Bingham, Dawson, \& Shapiro}]{bingham2002}
Bingham, R., Dawson, J.~M., \& Shapiro, V.~D. 2002, Journal of Plasma Physics,
  68, 161–172

\bibitem[{{Birdsall} \& {Langdon}(1991)}]{Birdsall_Langdon1991}
{Birdsall}, C.~K. \& {Langdon}, A.~B. 1991, {Plasma Physics via Computer
  Simulation}

\bibitem[{Brackbill {et~al.}(1984)Brackbill, Forslund, Quest, \&
  Winske}]{Brackbill1984}
Brackbill, J.~U., Forslund, D.~W., Quest, K.~B., \& Winske, D. 1984, The
  Physics of Fluids, 27, 2682

\bibitem[{Broiles {et~al.}(2016)Broiles, Burch, Chae, Clark, Cravens, Eriksson,
  Fuselier, Frahm, Gasc, Goldstein, Henri, Koenders, Livadiotis, Mandt,
  Mokashi, Nemeth, Odelstad, Rubin, \& Samara}]{Broiles2016}
Broiles, T.~W., Burch, J.~L., Chae, K., {et~al.} 2016, Monthly Notices of the
  Royal Astronomical Society, 462, S312

\bibitem[{{Brunetti} {et~al.}(2000){Brunetti}, {Califano}, \&
  {Pegoraro}}]{Brunetti2000}
{Brunetti}, M., {Califano}, F., \& {Pegoraro}, F. 2000, \pre, 62, 4109

\bibitem[{Bécoulet {et~al.}(2011)Bécoulet, Hoang, Artaud, Bae, Belo,
  Berger-By, Bernard, Cara, Cardinali, Castaldo, Ceccuzzi, Cesario, Cho,
  Decker, Delpech, Do, Ekedahl, Garcia, Garibaldi, Goniche, Guilhem,
  Hamlyn-Harris, Hillairet, Huang, Imbeaux, Jia, Kazarian, Kim, Lausenaz,
  Litaudon, Maggiora, Magne, Marfisi, Meschino, Milanesio, Mirizzi, Mollard,
  Namkung, Pajewski, Panaccione, Park, Park, Peysson, Saille, Samaille,
  Schettini, Schneider, Sharma, Tuccillo, Tudisco, Vecchi, Villari, Vulliez,
  Wu, Yang, \& Zeng}]{Becoulet2011}
Bécoulet, A., Hoang, G., Artaud, J., {et~al.} 2011, Fusion Engineering and
  Design, 86, 490 , proceedings of the 26th Symposium of Fusion Technology
  (SOFT-26)

\bibitem[{Cairns \& McMillan(2005)}]{Cairns2005}
Cairns, I.~H. \& McMillan, B.~F. 2005, Physics of Plasmas, 12, 102110

\bibitem[{Chen \& Birdsall(1983)}]{Chen1983}
Chen, Y. \& Birdsall, C.~K. 1983, The Physics of Fluids, 26, 180

\bibitem[{{Daughton}(1998)}]{Daughton1998}
{Daughton}, W. 1998, \jgr, 103, 29429

\bibitem[{{Daughton}(1999)}]{Daughton1999}
{Daughton}, W. 1999, \jgr, 104, 28701

\bibitem[{Daughton(2003)}]{daughton2003}
Daughton, W. 2003, Physics of Plasmas, 10, 3103

\bibitem[{Davidson(1978)}]{Davidson1978}
Davidson, R.~C. 1978, The Physics of Fluids, 21, 1375

\bibitem[{Davidson {et~al.}(1977)Davidson, Gladd, Wu, \& Huba}]{Davidson1977}
Davidson, R.~C., Gladd, N.~T., Wu, C.~S., \& Huba, J.~D. 1977, The Physics of
  Fluids, 20, 301

\bibitem[{Derouillat {et~al.}(2018)Derouillat, Beck, Pérez, Vinci,
  Chiaramello, Grassi, Flé, Bouchard, Plotnikov, Aunai, Dargent, Riconda, \&
  Grech}]{Derouillat2018}
Derouillat, J., Beck, A., Pérez, F., {et~al.} 2018, Computer Physics
  Communications, 222, 351

\bibitem[{Gary(1993)}]{Gary1993}
Gary, S.~P. 1993, {Theory of Space Plasma Microinstabilities}, Cambridge
  Atmospheric and Space Science Series (Cambridge University Press)

\bibitem[{Gary \& Sanderson(1978)}]{Gary1978}
Gary, S.~P. \& Sanderson, J.~J. 1978, The Physics of Fluids, 21, 1181

\bibitem[{Gary \& Sgro(1990)}]{Gary1990}
Gary, S.~P. \& Sgro, A.~G. 1990, Geophysical Research Letters, 17, 909

\bibitem[{Gershman {et~al.}(2015)Gershman, Raines, Slavin, Zurbuchen, Sundberg,
  Boardsen, Anderson, Korth, \& Solomon}]{Gershman2015}
Gershman, D.~J., Raines, J.~M., Slavin, J.~A., {et~al.} 2015, Journal of
  Geophysical Research: Space Physics, 120, 4354

\bibitem[{{Goldstein} {et~al.}(2019){Goldstein}, {Burch}, {Llera}, {Mokashi},
  {Nilsson}, {Dokgo}, {Eriksson}, {Odelstad}, \& {Richter}}]{Goldstein2019}
{Goldstein}, R., {Burch}, J.~L., {Llera}, K., {et~al.} 2019, Astronomy and
  Astrophysics, 630, A40

\bibitem[{Graham {et~al.}(2019)Graham, Khotyaintsev, Norgren, Vaivads, André,
  Drake, Egedal, Zhou, Le~Contel, Webster, Lavraud, Kacem, Génot, Jacquey,
  Rager, Gershman, Burch, \& Ergun}]{Graham2019}
Graham, D.~B., Khotyaintsev, Y.~V., Norgren, C., {et~al.} 2019, Journal of
  Geophysical Research: Space Physics, 124, 8727

\bibitem[{Graham {et~al.}(2017)Graham, Khotyaintsev, Norgren, Vaivads, André,
  Toledo-Redondo, Lindqvist, Marklund, Ergun, Paterson, Gershman, Giles,
  Pollock, Dorelli, Avanov, Lavraud, Saito, Magnes, Russell, Strangeway,
  Torbert, \& Burch}]{Graham2017}
Graham, D.~B., Khotyaintsev, Y.~V., Norgren, C., {et~al.} 2017, Journal of
  Geophysical Research: Space Physics, 122, 517

\bibitem[{{Hoshino} {et~al.}(2001){Hoshino}, {Mukai}, {Terasawa}, \&
  {Shinohara}}]{Hoshino2001}
{Hoshino}, M., {Mukai}, T., {Terasawa}, T., \& {Shinohara}, I. 2001, \jgr, 106,
  25979

\bibitem[{Huba {et~al.}(1978)Huba, Gladd, \& Papadopoulos}]{Huba1978}
Huba, J., Gladd, T., \& Papadopoulos, K. 1978, Journal of Geophysical Research,
  83

\bibitem[{Karlsson {et~al.}(2017)Karlsson, Eriksson, Odelstad, André, Dickeli,
  Kullen, Lindqvist, Nilsson, \& Richter}]{Karlsson2017}
Karlsson, T., Eriksson, A.~I., Odelstad, E., {et~al.} 2017, Geophysical
  Research Letters, 44, 1641

\bibitem[{{Kasaba} {et~al.}(2020){Kasaba}, {Kojima}, {Moncuquet}, {Wahlund},
  {Yagitani}, {Sahraoui}, {Henri}, {Karlsson}, {Kasahara}, {Kumamoto},
  {Ishisaka}, {Issautier}, {Wattieaux}, {Imachi}, {Matsuda}, {Lichtenberger},
  \& {Usui}}]{kasaba2020}
{Kasaba}, Y., {Kojima}, H., {Moncuquet}, M., {et~al.} 2020, \ssr, 216, 65

\bibitem[{{Khotyaintsev} {et~al.}(2011){Khotyaintsev}, {Cully}, {Vaivads},
  {Andr{\'e}}, \& {Owen}}]{Khotyaintsev2011}
{Khotyaintsev}, Y.~V., {Cully}, C.~M., {Vaivads}, A., {Andr{\'e}}, M., \&
  {Owen}, C.~J. 2011, \prl, 106, 165001

\bibitem[{Krall \& Liewer(1971)}]{Krall1971}
Krall, N.~A. \& Liewer, P.~C. 1971, Phys. Rev. A, 4, 2094

\bibitem[{{Krall} \& {Trivelpiece}(1973)}]{Krall1973}
{Krall}, N.~A. \& {Trivelpiece}, A.~W. 1973, {Principles of plasma physics}
  ({McGraw-Hill})

\bibitem[{{Krasnoselskikh} {et~al.}(1985){Krasnoselskikh}, {Kruchina},
  {Volokitin}, \& {Thejappa}}]{Volodia1985}
{Krasnoselskikh}, V.~V., {Kruchina}, E.~N., {Volokitin}, A.~S., \& {Thejappa},
  G. 1985, \aap, 149, 323

\bibitem[{{Laming}(2001)}]{Laming2001}
{Laming}, J.~M. 2001, \apj, 563, 828

\bibitem[{Lapenta \& Brackbill(2002)}]{Lapenta2002}
Lapenta, G. \& Brackbill, J.~U. 2002, Physics of Plasmas, 9, 1544

\bibitem[{{Lapenta} {et~al.}(2003){Lapenta}, {Brackbill}, \&
  {Daughton}}]{Lapenta2003}
{Lapenta}, G., {Brackbill}, J.~U., \& {Daughton}, W.~S. 2003, Physics of
  Plasmas, 10, 1577

\bibitem[{Lapenta {et~al.}(2018)Lapenta, Pucci, Olshevsky, Servidio,
  Sorriso-Valvo, Newman, \& Goldman}]{Lapenta2018}
Lapenta, G., Pucci, F., Olshevsky, V., {et~al.} 2018, Journal of Plasma
  Physics, 84, 715840103

\bibitem[{Le~Contel {et~al.}(2017)Le~Contel, Nakamura, Breuillard, Argall,
  Graham, Fischer, Retinò, Berthomier, Pottelette, Mirioni, Chust, Wilder,
  Gershman, Varsani, Lindqvist, Khotyaintsev, Norgren, Ergun, Goodrich, Burch,
  Torbert, Needell, Chutter, Rau, Dors, Russell, Magnes, Strangeway, Bromund,
  Wei, Plaschke, Anderson, Le, Moore, Giles, Paterson, Pollock, Dorelli,
  Avanov, Saito, Lavraud, Fuselier, Mauk, Cohen, Turner, Fennell, Leonard, \&
  Jaynes}]{LeContel2017}
Le~Contel, O., Nakamura, R., Breuillard, H., {et~al.} 2017, Journal of
  Geophysical Research: Space Physics, 122, 12,236

\bibitem[{Markidis {et~al.}(2010)Markidis, Lapenta, \&
  Rizwan-uddin}]{Markidis2010}
Markidis, S., Lapenta, G., \& Rizwan-uddin. 2010, Mathematics and Computers in
  Simulation, 80, 1509

\bibitem[{McBride \& Ott(1972)}]{McBride1972_beta}
McBride, J. \& Ott, E. 1972, Physics Letters A, 39, 363

\bibitem[{McBride {et~al.}(1972)McBride, Ott, Boris, \& Orens}]{McBride1972}
McBride, J.~B., Ott, E., Boris, J.~P., \& Orens, J.~H. 1972, The Physics of
  Fluids, 15, 2367

\bibitem[{{McClements} {et~al.}(1993){McClements}, {Bingham}, {Su}, {Dawson},
  \& {Spicer}}]{McClements1993}
{McClements}, K.~G., {Bingham}, R., {Su}, J.~J., {Dawson}, J.~M., \& {Spicer},
  D.~S. 1993, \apj, 409, 465

\bibitem[{McMillan(2020)}]{McMillan2020}
McMillan, B.~F. 2020, Physics of Plasmas, 27, 052106

\bibitem[{{Milillo} {et~al.}(2020){Milillo}, {Fujimoto}, {Murakami},
  {Benkhoff}, {Zender}, {Aizawa}, {D{\'o}sa}, {Griton}, {Heyner}, {Ho},
  {Imber}, {Jia}, {Karlsson}, {Killen}, {Laurenza}, {Lindsay},
  {McKenna-Lawlor}, {Mura}, {Raines}, {Rothery}, {Andr{\'e}}, {Baumjohann},
  {Berezhnoy}, {Bourdin}, {Bunce}, {Califano}, {Deca}, {de{\^A} la Fuente},
  {Dong}, {Grava}, {Fatemi}, {Henri}, {Ivanovski}, {Jackson}, {James},
  {Kallio}, {Kasaba}, {Kilpua}, {Kobayashi}, {Langlais}, {Leblanc}, {Lhotka},
  {Mangano}, {Martindale}, {Massetti}, {Masters}, {Morooka}, {Narita},
  {Oliveira}, {Odstrcil}, {Orsini}, {Pelizzo}, {Plainaki}, {Plaschke},
  {Sahraoui}, {Seki}, {Slavin}, {Vainio}, {Wurz}, {Barabash}, {Carr},
  {Delcourt}, {Glassmeier}, {Grande}, {Hirahara}, {Huovelin}, {Korablev},
  {Kojima}, {Lichtenegger}, {Livi}, {Matsuoka}, {Moissl}, {Moncuquet},
  {Muinonen}, {Qu{\`e}merais}, {Saito}, {Yagitani}, {Yoshikawa}, \&
  {Wahlund}}]{Milillo2020}
{Milillo}, A., {Fujimoto}, M., {Murakami}, G., {et~al.} 2020, \ssr, 216, 93

\bibitem[{Norgren {et~al.}(2012)Norgren, Vaivads, Khotyaintsev, \&
  Andr\'e}]{Norgren2012}
Norgren, C., Vaivads, A., Khotyaintsev, Y.~V., \& Andr\'e, M. 2012, Phys. Rev.
  Lett., 109, 055001

\bibitem[{{Ogilvie} {et~al.}(1974){Ogilvie}, {Scudder}, {Hartle}, {Siscoe},
  {Bridge}, {Lazarus}, {Asbridge}, {Bame}, \& {Yeates}}]{Ogilvie1974}
{Ogilvie}, K.~W., {Scudder}, J.~D., {Hartle}, R.~E., {et~al.} 1974, Science,
  185, 145

\bibitem[{Ott {et~al.}(1972)Ott, McBride, Orens, \& Boris}]{Ott1972}
Ott, E., McBride, J.~B., Orens, J.~H., \& Boris, J.~P. 1972, Phys. Rev. Lett.,
  28, 88

\bibitem[{Pericoli-Ridolfini {et~al.}(1999)Pericoli-Ridolfini, Barbato, Cirant,
  Kroegler, Panaccione, Podda, Alladio, Angelini, Apicella, Apruzzese,
  Bertalot, Bertocchi, Borra, Bracco, Bruschi, Buceti, Buratti, Cardinali,
  Centioli, Cesario, Ciattaglia, Cocilovo, Crisanti, De~Angelis, De~Marco,
  Esposito, Frigione, Gabellieri, Gatti, Giovannozzi, Gourlan, Granucci,
  Gravanti, Grolli, Imparato, Leigheb, Lovisetto, Maffia, Maddaluno, Marinucci,
  Mazzitelli, Micozzi, Mirizzi, Nowak, Orsitto, Pacella, Panella, Pieroni,
  Righetti, Romanelli, Santini, Sassi, Segre, Simonetto, Sozzi, Sternini,
  Tudisco, Vitale, Vlad, Tartoni, Tilia, Tuccillo, Zanza, Zerbini, \&
  Zonca}]{Pericoli-Ridolfini1999}
Pericoli-Ridolfini, V., Barbato, E., Cirant, S., {et~al.} 1999, Phys. Rev.
  Lett., 82, 93

\bibitem[{Pritchett {et~al.}(1996)Pritchett, Coroniti, \&
  Decyk}]{Pritchett1996}
Pritchett, P.~L., Coroniti, F.~V., \& Decyk, V.~K. 1996, Journal of Geophysical
  Research: Space Physics, 101, 27413

\bibitem[{Pu {et~al.}(1981)Pu, Quest, Kivelson, \& Tu}]{Pu1981}
Pu, Z.-Y., Quest, K.~B., Kivelson, M.~G., \& Tu, C.-Y. 1981, Journal of
  Geophysical Research: Space Physics, 86, 8919

\bibitem[{{Reiniusson} {et~al.}(2006){Reiniusson}, {Stenberg}, {Norqvist},
  {Eriksson}, \& {R{\"o}nnmark}}]{Reiniusson2006}
{Reiniusson}, A., {Stenberg}, G., {Norqvist}, P., {Eriksson}, A.~I., \&
  {R{\"o}nnmark}, K. 2006, Annales Geophysicae, 24, 367

\bibitem[{Retinò {et~al.}(2008)Retinò, Nakamura, Vaivads, Khotyaintsev,
  Hayakawa, Tanaka, Kasahara, Fujimoto, Shinohara, Eastwood, André,
  Baumjohann, Daly, Kronberg, \& Cornilleau-Wehrlin}]{Retino2008}
Retinò, A., Nakamura, R., Vaivads, A., {et~al.} 2008, Journal of Geophysical
  Research: Space Physics, 113

\bibitem[{Rigby {et~al.}(2018)Rigby, Cruz, Albertazzi, Bamford, Bell, Cross,
  Fraschetti, Graham, Hara, Kozlowski, Kuramitsu, Lamb, Lebedev, Marques,
  Miniati, Morita, Oliver, Reville, Sakawa, \& Gregori}]{rigby2018}
Rigby, A., Cruz, F., Albertazzi, B., {et~al.} 2018, Nature Physics, 14

\bibitem[{{Russell} {et~al.}(1988){Russell}, {Baker}, \&
  {Slavin}}]{Russell1988}
{Russell}, C.~T., {Baker}, D.~N., \& {Slavin}, J.~A. 1988, {The magnetosphere
  of Mercury.}, ed. F.~{Vilas}, C.~R. {Chapman}, \& M.~S. {Matthews}, 514--561

\bibitem[{Sagdeev {et~al.}(1990)Sagdeev, Shapiro, Shevchenko, Zacharov,
  Király, Szegó, Nagy, \& Grard}]{Sagdeev1990}
Sagdeev, R.~Z., Shapiro, V.~D., Shevchenko, V.~I., {et~al.} 1990, Geophysical
  Research Letters, 17, 893

\bibitem[{{Saito} {et~al.}(2010){Saito}, {Sauvaud}, {Hirahara}, {Barabash},
  {Delcourt}, {Takashima}, {Asamura}, \& {BepiColombo MMO/MPPE
  Team}}]{Saito2010}
{Saito}, Y., {Sauvaud}, J.~A., {Hirahara}, M., {et~al.} 2010, \planss, 58, 182

\bibitem[{Scarf {et~al.}(1980)Scarf, Taylor, Russell, \& Elphic}]{Scarf1980}
Scarf, F.~L., Taylor, W. W.~L., Russell, C.~T., \& Elphic, R.~C. 1980, Journal
  of Geophysical Research: Space Physics, 85, 7599

\bibitem[{Shapiro {et~al.}(1999)Shapiro, Bingham, Dawson, Dobe, Kellett, \&
  Mendis}]{Shapiro1999}
Shapiro, V.~D., Bingham, R., Dawson, J.~M., {et~al.} 1999, Journal of
  Geophysical Research: Space Physics, 104, 2537

\bibitem[{{Shapiro} {et~al.}(1995){Shapiro}, {Szeg{\"o}}, {Ride}, {Nagy}, \&
  {Shevchenko}}]{Shapiro1995}
{Shapiro}, V.~D., {Szeg{\"o}}, K., {Ride}, S.~K., {Nagy}, A.~F., \&
  {Shevchenko}, V.~I. 1995, \jgr, 100, 21289

\bibitem[{{Shinohara} \& {Hoshino}(1999)}]{Shinohara1999}
{Shinohara}, I. \& {Hoshino}, M. 1999, Advances in Space Research, 24, 43

\bibitem[{{Singh} {et~al.}(1996){Singh}, {Al-Sharaeh}, {Abdelrazek}, {Leung},
  \& {Wells}}]{Singh1996}
{Singh}, N., {Al-Sharaeh}, S., {Abdelrazek}, A., {Leung}, W.~C., \& {Wells},
  B.~E. 1996, \grl, 23, 3663

\bibitem[{{Singh} {et~al.}(1998){Singh}, {Wells}, {Abdelrazek}, {Al-Sharaeh},
  \& {Leung}}]{Singh1998}
{Singh}, N., {Wells}, B.~E., {Abdelrazek}, A., {Al-Sharaeh}, S., \& {Leung},
  W.~C. 1998, Journal of Geophysics Research, 103, 9333

\bibitem[{Slavin {et~al.}(2008)Slavin, Acu{\~n}a, Anderson, Baker, Benna,
  Gloeckler, Gold, Ho, Killen, Korth, Krimigis, McNutt, Nittler, Raines,
  Schriver, Solomon, Starr, Tr{\'a}vn{\'\i}{\v c}ek, \& Zurbuchen}]{Slavin2008}
Slavin, J.~A., Acu{\~n}a, M.~H., Anderson, B.~J., {et~al.} 2008, 321, 85

\bibitem[{{Solomon} {et~al.}(2007){Solomon}, {McNutt}, {Gold}, \&
  {Domingue}}]{Solomon2007}
{Solomon}, S.~C., {McNutt}, R.~L., {Gold}, R.~E., \& {Domingue}, D.~L. 2007,
  Space Science Reviews, 131, 3

\bibitem[{{Tang} {et~al.}(2020){Tang}, {Li}, {Graham}, {Wang}, {Khotyaintsev},
  {Le}, {Giles}, {Lindqvist}, {Ergun}, \& {Burch}}]{Tang2020}
{Tang}, B.~B., {Li}, W.~Y., {Graham}, D.~B., {et~al.} 2020, \grl, 47, e89880

\bibitem[{{Vaivads} {et~al.}(2004){Vaivads}, {Andr{\'e}}, {Buchert}, {Wahlund},
  {Fazakerley}, \& {Cornilleau-Wehrlin}}]{Vaivads2004}
{Vaivads}, A., {Andr{\'e}}, M., {Buchert}, S.~C., {et~al.} 2004, \grl, 31,
  L03804

\bibitem[{{Walker} {et~al.}(2008){Walker}, {Balikhin}, {Alleyne}, {Hobara},
  {Andr{\'e}}, \& {Dunlop}}]{Walker2008}
{Walker}, S.~N., {Balikhin}, M.~A., {Alleyne}, H. S. C.~K., {et~al.} 2008,
  Annales Geophysicae, 26, 699

\bibitem[{{Wilson} {et~al.}(2013){Wilson}, {Koval}, {Szabo}, {Breneman},
  {Cattell}, {Goetz}, {Kellogg}, {Kersten}, {Kasper}, {Maruca}, \&
  {Pulupa}}]{Wilson2013}
{Wilson}, L.~B., {Koval}, A., {Szabo}, A., {et~al.} 2013, Journal of
  Geophysical Research (Space Physics), 118, 5

\bibitem[{{Yoo} {et~al.}(2020){Yoo}, {Ji}, {Ambat}, {Wang}, {Ji}, {Lo}, {Li},
  {Ren}, {Jara-Almonte}, {Chen}, {Fox}, {Yamada}, {Alt}, \&
  {Goodman}}]{Yoo2020}
{Yoo}, J., {Ji}, J.-Y., {Ambat}, M.~V., {et~al.} 2020, \grl, 47, e87192

\bibitem[{{Zacharegkas} {et~al.}(2016){Zacharegkas}, {Isliker}, \&
  {Vlahos}}]{Zacharegkas2016}
{Zacharegkas}, G., {Isliker}, H., \& {Vlahos}, L. 2016, Physics of Plasmas, 23,
  112119

\bibitem[{Zhang \& Matsumoto(1998)}]{Zhang1998}
Zhang, Y. \& Matsumoto, H. 1998, Journal of Geophysical Research: Space
  Physics, 103, 20561

\bibitem[{Zhou {et~al.}(2009)Zhou, Deng, Li, Pang, Vaivads, Rème, Lucek, Fu,
  Lin, Yuan, \& Wang}]{Zhou2009}
Zhou, M., Deng, X.~H., Li, S.~Y., {et~al.} 2009, Journal of Geophysical
  Research: Space Physics, 114

\bibitem[{Zhou {et~al.}(2014)Zhou, Li, Deng, Huang, Pang, Yuan, Xu, \&
  Tang}]{Zhou2014}
Zhou, M., Li, H., Deng, X., {et~al.} 2014, Journal of Geophysical Research:
  Space Physics, 119, 8228

\end{thebibliography}

\end{document}